\documentclass[submission,copyright]{eptcs}
\providecommand{\event}{ICE 2019} 
\usepackage{amsthm}
\usepackage{amsfonts}
\usepackage{amsmath}
\usepackage{amssymb}
\usepackage{amsthm}
\usepackage{shuffle}

\DeclareFontFamily{U}{matha}{}
\DeclareFontShape{U}{matha}{m}{n}{
  <-5.5>    matha5
  <5.5-6.5> matha6 
  <6.5-7.5> matha7
  <7.5-8.5> matha8
  <8.5-9.5> matha9
  <9.5-11>  matha10
  <11->     matha12
}{}
\DeclareSymbolFont{matha}{U}{matha}{m}{n}
\DeclareFontSubstitution{U}{matha}{m}{n}
\DeclareFontFamily{U}{mathx}{\hyphenchar\font45}
\DeclareFontShape{U}{mathx}{m}{n}{<-> mathx10}{}
\DeclareSymbolFont{mathx}{U}{mathx}{m}{n}
\DeclareFontSubstitution{U}{mathx}{m}{n}

\DeclareMathDelimiter{\ldbrack}{4}{matha}{"76}{mathx}{"30}
\DeclareMathDelimiter{\rdbrack}{5}{matha}{"77}{mathx}{"38}

\usepackage{bbm}
\usepackage{color}
\usepackage[all]{xy}
\usepackage{tikz}
\usetikzlibrary{arrows,automata}
\tikzset{initial text={}}
\usepackage{comment}
\usepackage{pgfgantt}
\usepackage{cleveref}

\usepackage[nomargin,multiuser,draft,inline]{fixme}
\usepackage{todonotes}
\fxusetheme{color}
\FXRegisterAuthor{eM}{em}{{\color{orange} \underline{eM} }\todo{!}}

\usetikzlibrary{arrows.meta,shapes.geometric}
\tikzset{%
  >={Latex[width=2mm,length=2mm]},
 base/.style = {rectangle, rounded corners, draw=black,      
                           text centered, font=\sffamily},              
          folk/.style={regular polygon, draw, regular polygon sides=4,inner sep=0.3pt,minimum size=1pt},
          start/.style = {circle, draw,inner sep=2pt,minimum size=1pt},
          end/.style = {circle, double, draw, fill=blue!30,inner sep=2pt,minimum size=1pt},
          branch/.style = {diamond, draw, inner sep=0.3pt,minimum size=1pt},
          basic/.style = {base,fill=blue!10},
}

\tikzset{
    ncbar angle/.initial=90,
    ncbar/.style={
        to path=(\tikztostart)
        -- ($(\tikztostart)!#1!\pgfkeysvalueof{/tikz/ncbar angle}:(\tikztotarget)$)
        -- ($(\tikztotarget)!($(\tikztostart)!#1!\pgfkeysvalueof{/tikz/ncbar angle}:(\tikztotarget)$)!\pgfkeysvalueof{/tikz/ncbar angle}:(\tikztostart)$)
        -- (\tikztotarget)
    },
    ncbar/.default=0.5cm,
}

\tikzset{square left brace/.style={ncbar=0.2cm}}
\tikzset{square right brace/.style={ncbar=-0.2cm}}

\usepackage{caption}
\usepackage{wrapfig}
\usepackage{setspace}
\usepackage{algorithm} 
\usepackage{algorithmic} 
\usepackage{mathtools}
\usepackage{makeidx}
\makeindex
\usepackage{hyperref}

\theoremstyle{plain}
\newtheorem{theorem}{Theorem}[section]

\theoremstyle{definition}
\newtheorem{definition}[theorem]{Definition}

\newcommand{\quo}[1]{\lq\lq {#1}\rq\rq}

\title{Interface Automata for Choreographies}
\author{
  Hao Zeng
  \institute{Department of Informatics\\
    University of Leicester\\
    Leicester, UK}
  \email{hz110@le.ac.uk}
  \and
  Alexander Kurz
  \institute{Chapman University\\
    California, USA\\\ }
  \email{akurz@chapman.edu}
  \and
  Emilio Tuosto
  \institute{Gran Sasso Science Institute, IT and
    \\ Department of Informatics\\University of Leicester, UK}
  \email{emilio.tuosto@gssi.it}
}
\def\titlerunning{Interface Automata for Choreographies}
\def\authorrunning{}
\def\authorrunning{H. Zeng, A. Kurz \& E. Tuosto}
\begin{document}
\maketitle
\begin{abstract}
  Choreographic approaches to message-passing applications can be
  regarded as an instance of the model-driven development principles.
  Choreographies specify interactions among distributed participants
  coordinating among themselves with message-passing at two levels
  of abstractions.
  A \emph{global view} of the application is specified with a model
  that abstracts away from asynchrony while a \emph{local view} of the
  application specifies the communication pattern of each participant.
  Noteworthy, the latter view can typically be algorithmically obtained
  by \emph{projection} of the global view.
  A crucial element of this approach is to verify the so-called
  \emph{well-formed} conditions on global views so that its
  projections realise a sound communication protocol.
  We introduce a novel local model, group interface automata, to
  represent the local view of choreographies and propose a new method
  to verify the well-formedness of global choreographies. We rely on a
  recently proposed semantics of global views formalised in terms of
  pomsets.
\end{abstract}

\newcommand{\thegg}{
\begin{wrapfigure}[18]{c}[0pt]{0.4\textwidth}
  \captionsetup{justification=centering}
  \vspace{-.3cm}
  \begin{center}
    \begin{tikzpicture}[node distance=0.9cm, every node/.style={fill=white, font=\sffamily}, align=center]
      \node (0)             [start]              {};
      \node (1)     [basic, below of=0]          {\scriptsize B $\xrightarrow{\text{request}}$ S};
      \node (2)      [branch, below of=1]   {\tiny $+$};
      \node at (-1.3, -2.5) (3)     [basic]   {\scriptsize S $\xrightarrow{\text{offer}}$ B};
      \node (4)      [basic, below of=3]  {\scriptsize B $\xrightarrow{\text{pay}}$ S};
      \node (5)      [basic, below of=4]      {\scriptsize S $\xrightarrow{\text{deliveryInfo}}$ H};
      \node (6)       [basic, below of=5]                   {\scriptsize H $\xrightarrow{\text{delivery}}$ B};
      \node at (1.3, -2.5) (7)     [basic]   {\scriptsize S $\xrightarrow{\text{notinStock}}$ B};
      \node (8)      [basic, below of=7]  {\scriptsize S $\xrightarrow{\text{noInfo}}$ H};
      \node at (0, -5.9) (9)      [branch]   {\tiny $+$};
      \node at (0, -6.6) (10)    [end]              {};
      \draw[-{stealth}]      (0) -- (1);
      \draw[-{stealth}]      (1) -- (2);
      \draw[-{stealth}]      (2) -| (3);
      \draw[-{stealth}]      (3) -- (4);
      \draw[-{stealth}]      (4) -- (5);
      \draw[-{stealth}]      (5) -- (6);
      \draw[-{stealth}]      (2) -| (7);
      \draw[-{stealth}]      (7) -- (8);
      \draw[-{stealth}]      (6) |- (9);
      \draw[-{stealth}]      (8) |- (9);
      \draw[-{stealth}]      (9) -- (10);
    \end{tikzpicture}
  \end{center}
  \caption{A simple g-choreography} \label{fig:gg}
\end{wrapfigure}
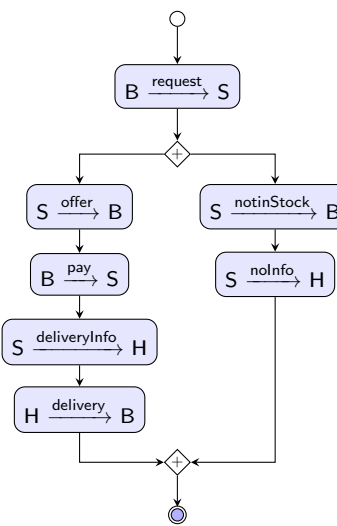
}

\section{Introduction}


Nowadays distributed applications are widely used in our daily life,
ranging from online payments via social communications and web
services to multi-core computing.
The engineering of distributed systems inevitably entwines with
communication among components, which is a key element for realising
distributed coordination mechanism.
The formalisation of coordination protocols becomes an important
challenge.
Indeed the use of abstract models to tame the complexity of
distributed applications
\thegg
is becoming commonplace also in industrial
context~\cite{boner18}.
To this end, \emph{choreographies}~\cite{kavantzas2005web} have been
proposed as a methodology to facilitate the coordination of
distributed components.
Roughly speaking, a choreography is made of two elements, the
\emph{global view} and the \emph{local view}~\cite{kavantzas2005web}.
The former
specifies the interactions among distributed components in terms of a
so-called application level protocol, where the behaviour of the
system is described in terms of the relations among the (\emph{role}
of each) component.  The local view yields a more concrete
specification whereby the behaviour of each component \quo{in
  isolation} is derived from its role in the global view; at this
level of abstraction, components can be thought of as autonomous
agents enacting a role described in the global view, \emph{regardless of}
the other components.

In this paper, we are interested in message-passing systems and we
adopt a variant of \emph{g-choreographies} (after \emph{global
  choreographies})~\cite{gt16,gt17} as a
formalisation of global views of this application domain.
A main motivation to borrow g-choreographies is that they are rather
expressive (see the discussion in~\cite{gt16,gt17}) and
have an abstract semantics based on pomsets.
For instance, g-choreographies abstract away from asynchrony
(interactions are atomic and not specified in terms of send/receive
actions of participants).
In this way, g-choreographies simplify the description of the system
by relegating the complexity of asynchronous communication in the
local views.
We illustrate the model through an online shopping scenario given in
Figure~\ref{fig:gg}.
Buyer \textsf{B} sends a \textsf{request} to
Seller \textsf{S}, asking for some goods.
After the reception of this message, \textsf{S} decides to send ($i$) an
\textsf{offer} back to \textsf{B} or ($ii$) to a \textsf{notinStock}
message to \textsf{B} depending on the availability of the goods.
In the first case, \textsf{B} sends a message \textsf{pay} to
\textsf{S} representing a payment.
After receiving the online payment, \textsf{S} sends the
\textsf{deliveryInfo} to a shipper \textsf{H}. Finally, \textsf{H} sends
a \textsf{delivery} message to \textsf{B}.
If the alternative ($ii$) is taken, \textsf{S} sends a \textsf{noInfo}
message to \textsf{H} notifying that this purchase is not successful.
Figure~\ref{fig:gg} gives a visual description of this g-choreography.
Note that the g-choreography of this simple protocol clearly specifies
where the choice made by \textsf{S} takes place and where it finishes.
Also, observe that the choice is locally made by \textsf{S} and propagated to
the other participants.

We opt for local views of choreographies describing the behaviour of each
participant in terms of send and receive actions.
More precisely, we introduce a class of interface
automata~\cite{de2001interface, de2005interface} and show how they can
be used to check the correctness of a model of global specifications.
In our formal framework, the local views of each participant are
obtained by projections from a g-choreography.
In fact, the relationship between g-choreography and local views can
be represented by following diagram:
\[
  \text{G-choreography} \xrightarrow{\textit{projection}}   \text{Interface automata} \xleftarrow {\textit{comply}}  \text{Local System}
\]
Here we neglect the \textit{comply} relation and focus only on the
\textit{projection} one.
In particular, we show how our interface automata can be used to check
the \emph{well-formedness} of g-choreographies.\footnote{
  As known from the literature, well-formedness of g-choreographies
  guarantees well-behaviour of local components projected from the
  global specification (see
  e.g.~\cite{hyc08,lty15,gt16,gt17})
  .
}
Technically, we define an internal product to the class of our
interface automata and show that checking some conditions on these
products is equivalent to checking for the well-formedness of
g-choreographies.

We extend interface automata~\cite{de2001interface, de2005interface}
by defining \emph{group interface automata}.
The group interface automaton $\mathcal{I}_\textsf{B}$ for the local view of
buyer \textsf{B} of Figure~\ref{fig:gg} is
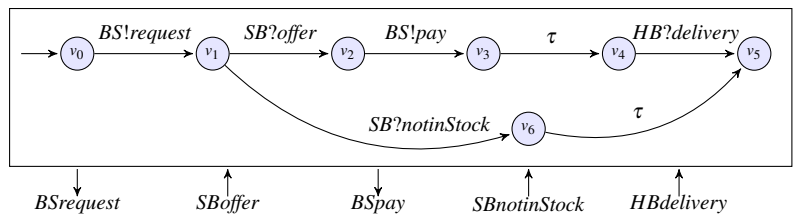
\[
  \begin{tikzpicture}[shorten >=1pt,node distance=1.8cm,>=stealth',auto, every state/.style={thin,fill=blue!10}]     
    \draw (-0.9,-1.5) rectangle (9.5,0.6);                                    
    \node at (0,-2.0) {\scriptsize $BS\textit{request}$}; 
    \node at (2,-2.0) {\scriptsize $SB\textit{offer}$};   
    \node at (4,-2.0) {\scriptsize $BS\textit{pay}$};
    \node at (6,-2.0) {\scriptsize $SB\textit{notinStock}$};
    \node at (8,-2.0) {\scriptsize $HB\textit{delivery}$};
    \draw  [<-] (0,-1.9)--(0,-1.5);
    \draw  [->] (2,-1.9)--(2,-1.5); 
    \draw  [<-] (4,-1.9)--(4,-1.5);
    \draw  [->] (6,-1.9)--(6,-1.5); 
    \draw  [->] (8,-1.9)--(8,-1.5);           
    \node[state,initial,inner sep=2pt,minimum size=0pt]        (q_0)                           {\tiny $v_0$};  
    \node[state,inner sep=2pt,minimum size=0pt]  (q_1)  [right of=q_0]     {\tiny $v_1$}; 
    \node[state,inner sep=2pt,minimum size=0pt]  (q_2)  [right of=q_1]     {\tiny $v_2$};
    \node[state,inner sep=2pt,minimum size=0pt]  (q_3)  [right of=q_2]     {\tiny $v_3$};
    \node[state,inner sep=2pt,minimum size=0pt]  (q_4)  [right of=q_3]     {\tiny $v_4$};
    \node[state,inner sep=2pt,minimum size=0pt]  (q_5)  [right of=q_4]     {\tiny $v_5$};
    \node[state,inner sep=2pt,minimum size=0pt]  (q_6)  at (6,-1.0)     {\tiny $v_6$};
    
    \path[->]       
    (q_0)     edge    node{\scriptsize ${BS!\textit{request}}$}     (q_1)
    (q_1)     edge    node{\scriptsize ${SB?\textit{offer}}$}     (q_2)
    edge  [bend right]  node{\scriptsize ${SB?\textit{notinStock}}$}     (q_6)
    (q_2)     edge    node{\scriptsize ${BS!\textit{pay}}$}     (q_3)
    (q_3)     edge    node{\scriptsize ${\tau}$}     (q_4)
    (q_4)     edge    node{\scriptsize ${HB?\textit{delivery}}$}     (q_5)
    (q_6)     edge [bend right]   node{\scriptsize ${\tau}$}     (q_5);
  \end{tikzpicture}
\]
%

Roughly speaking, the rectangle specifies the scope of the machine, the down and up arrows on the edge of the rectangle describe the output and input interfaces respectively (for more details see  section~\ref{sectGIA}). 


\paragraph{Main contributions}
This paper is based on the choreographic framework presented
in~\cite{gt16,gt17}.
We elaborate on the pomset semantics for global specifications
(g-choreographies) given in~\cite{gt16,gt17} and
formalising the interplay between global and local specifications.
More precisely, we reduce the notion of \emph{well-formedness} of
g-choreographies given by Guanciale and Tuosto to the analysis of (an
extension of) interface automata~\cite{de2001interface}.
The notion of well-formedness identifies a sufficient condition to
guarantee that the asynchronous execution of the projections of a
g-choreographies is sound, that is the execution is deadlock-free and
without orphan messages or unspecificed
receptions~\cite{brand1983communicating}.

The main contribution of this paper is the
reduction of well-formedness of g-choreographies to the absence of
\emph{error-states} in a variant of interface automata.
More precisely we show that a g-choreography $\textsf{G}$ is
well-formedness if, and only if, the group interface automaton
consisting of $\textsf{G}$'s projections does not contain \emph{error-states}.
This result  requires some technical contributions.

Firstly, we extend interface automata to \emph{group interface
  automata} (GIA) to represent local views of global choreographies.
Secondly, we define a product operation on GIA that allows us to
identify a class of configurations, dubbed \emph{error states}
that may spoil communications.
The identification of error-states is based on the notion of
\emph{removable} internal transition of GIA
that we define here.

\paragraph{Related work}
We review the choreographic models that are closest to the model presented in this paper.
%
%
  \emph{Global graphs} were introduced
  in~\cite{denielou2012multiparty} as graphical interpretation of
  \emph{global types}~\cite{hyc08} and then refined
  in~\cite{gt16,gt17} as g-choreograhies.
  This paper uses a variant of g-choreographies
  (cf. Definition~\ref{semanticOfGGSimple}) and their pomset semantics as
  the global view of choreographies.
  The variant disregards the requirements of well-sequencedness imposed in~\cite{gt16,gt17},
  uses
  well-forkedness in~\cite{gt16,gt17} and a relaxed notion of well-branchedness
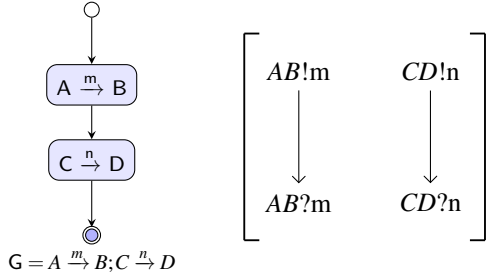
\begin{wrapfigure}[10]{l}[0pt]{0.5\textwidth}
\captionsetup{justification=centering}
\begin{minipage}{0.24\textwidth}
\centering
\begin{tikzpicture}[node distance=1cm,
    every node/.style={fill=white, font=\sffamily}, align=center]
  \node (0)             [start]              {};
  \node (1)     [basic, below of=0]          {\scriptsize A $\xrightarrow{\text{m}}$ B};
  \node (2)      [basic, below of=1]   {\scriptsize C $\xrightarrow{\text{n}}$ D};
  \node (3)    [end, below of=2]              {};
  \draw[-{stealth}]      (0) -- (1);
  \draw[-{stealth}]      (1) -- (2);
  \draw[-{stealth}]      (2) -- (3);
\end{tikzpicture}
\centering
\\
\scriptsize $\textsf{G}=A\xrightarrow{m}B ;C\xrightarrow{n}D$ 
\end{minipage}
\begin{minipage}{0.24\textwidth}
\begin{tikzpicture}[node distance=1.75cm]
\draw [line width=0.25mm] (-1.6,0) -- (-1.6,-2.75)  (1.6,0) -- (1.6,-2.75);
\draw [line width=0.25mm] (-1.6,0) -- (-1.4,0)  (1.6,0) -- (1.4,0);
\draw [line width=0.25mm] (-1.6,-2.75) -- (-1.4,-2.75)  (1.6,-2.75) -- (1.4,-2.75);

\node at (-0.875, -0.5)   (a) {\small \textit{AB}!m};
\node[right of=a]   (b) {\small \textit{CD}!n};
\node[below of=a]   (c) {\small \textit{AB}?m};
\node[right of=c]   (d) {\small \textit{CD}?n};

\draw[arrows={-angle 90}] (a)--(c);
\draw[arrows={-angle 90}] (b)--(d);
\end{tikzpicture}
\end{minipage}
\end{wrapfigure}
for simplicity.
%
For instance, a sequential composition $\textsf{G}$ with its pomset
semantics in our variant are shown on the left figure.
According to~\cite{gt16,gt17}, the semantics of
$\textsf{G}$ is undefined since $\textsf{G}$ violates
well-sequencedness due to the lack of causal dependencies among the
participants of the two interactions.
Instead, our variant gives $\textit{G}$ the semantics represented by
the pomset shown in the right-hand-side of the figure, which
simply allows $A\xrightarrow{m}B$ and $C\xrightarrow{n}D$ to run
concurrently.
Therefore, the semantics of $\textsf{G}$ is equivalent to the
semantics of the parallel composition
$A\xrightarrow{m}B \mid
C\xrightarrow{n}D$. 
%
%

Next, we review the local models. \emph{Interface automata}~\cite{de2001interface, de2005interface} 
are a class of synchronous local models to support component-based design and
verification in software engineering. Two composable interface automata are allowed to interact via a product that respects their interfaces. 
An error state $(v,u)$  in a product automaton consists of a state $v$ in which a send in the shared output interface will not be consumed from the corresponding state $u$ onwards. 
However, the notion of error states in interface automata is too strong for interface automata as local views of global choreographies.
\emph{Communicating finite state machines (CFSM)}~\cite{brand1983communicating} are a convient setting to analyse
choreographies from a local point of view. However, CFSM do not have interfaces and product operations.   
In order to combine advantages from interface automata and CFSM, we propose \emph{group interface automata} as an extension to interface automata.
Group interface automata strengthen the sender and receiver on interfaces, add a notion of special internal $\tau$-transitions, and redefine products and error states. These changes allow us to adopt interface automata to anaylse global choreographies (for details see sections~\ref{sectGIA} and~\ref{sectGIAbV}).



\section{Background}
This section summarises the main concepts we use in the paper. 
We adapt the definitions from \cite{gt16}. We write $\mathcal{P}$ for a set of participants and  $\mathcal{M}$ for a set of messages.

\begin{figure}[htbp]
\begin{minipage}{0.19\textwidth}
\centering
\begin{tikzpicture}[node distance=1cm,
    every node/.style={fill=white, font=\sffamily}, align=center]
  \node (0)             [start]              {};
  \node (1)    [end, below of=0]              {};
  \draw[-{stealth}]      (0) -- (1);
\end{tikzpicture}
\centering
\\  $(1)$ empty
\end{minipage}
\begin{minipage}{0.19\textwidth}
\centering
\begin{tikzpicture}[node distance=1cm,
    every node/.style={fill=white, font=\sffamily}, align=center]

  \node (0)             [start]              {};
  \node (1)     [basic, below of=0]          {\scriptsize A $\xrightarrow{\text{m}}$ B};
  \node (2)    [end, below of=1]              {};
  \draw[-{stealth}]      (0) -- (1);
  \draw[-{stealth}]     (1) -- (2);
\end{tikzpicture}
\centering
\\  $(2)$ interaction
\end{minipage}
\begin{minipage}{0.19\textwidth}
\centering
\begin{tikzpicture}[node distance=0.9cm,
    every node/.style={fill=white, font=\sffamily}, align=center]

  \node (0)             [start]              {};
  \node (1)     [basic, below of=0]          {\scriptsize \textsf{G}};
  \node (2)      [basic, below of=1]   {\scriptsize\textsf{G'}};
  \node (3)    [end, below of=2]              {};
  \draw[-{stealth}]      (0) -- (1);
  \draw[-{stealth}]     (1) -- (2);
  \draw[-{stealth}]      (2) -- (3);
\end{tikzpicture}
\centering
\\  $(3)$ sequential
\end{minipage}
\begin{minipage}{0.19\textwidth}
\centering
\begin{tikzpicture}[node distance=1cm,
    every node/.style={fill=white, font=\sffamily}, align=center]
  \node (0)             [start]              {};
  \node at (0, -0.8) (1)     [folk]          {\tiny $|$};
  \node (2)     [basic, below left of=1]          {\scriptsize \textsf{G}};
  \node (3)      [basic, below right of=1]   {\scriptsize \textsf{G'}};
  \node (4)    [folk, below right of=2]              {\tiny $|$};
  \node at (0, -3.0) (5)    [end]              {};
  \draw[-{stealth}]      (0) -- (1);
  \draw[-{stealth}]     (1) -| (2);
  \draw[-{stealth}]      (1) -| (3);
  \draw[-{stealth}]      (2) |- (4);
  \draw[-{stealth}]      (3) |- (4);
  \draw[-{stealth}]      (4) -- (5);
\end{tikzpicture}
\centering
\\  $(4)$ parallel
\end{minipage}
\begin{minipage}{0.19\textwidth}
\centering
\begin{tikzpicture}[node distance=1cm,
    every node/.style={fill=white, font=\sffamily}, align=center]
  \node (0)             [start]              {};
  \node at (0, -0.8) (1)     [branch]          {\tiny $+$};
  \node (2)     [basic, below left of=1]          {\scriptsize \textsf{G}};
  \node (3)      [basic, below right of=1]   {\scriptsize \textsf{G'}};
  \node (4)    [branch, below right of=2]              {\tiny $+$};
  \node at (0, -3.0) (5)    [end]              {};
  \draw[-{stealth}]      (0) -- (1);
  \draw[-{stealth}]     (1) -| (2);
  \draw[-{stealth}]      (1) -| (3);
  \draw[-{stealth}]      (2) |- (4);
  \draw[-{stealth}]      (3) |- (4);
  \draw[-{stealth}]      (4) -- (5);
\end{tikzpicture}
\centering
\\  $(5)$ branching
\end{minipage}
\caption{The visual notations of g-choreographies} \label{VisualNotaionOfGG}
\end{figure}
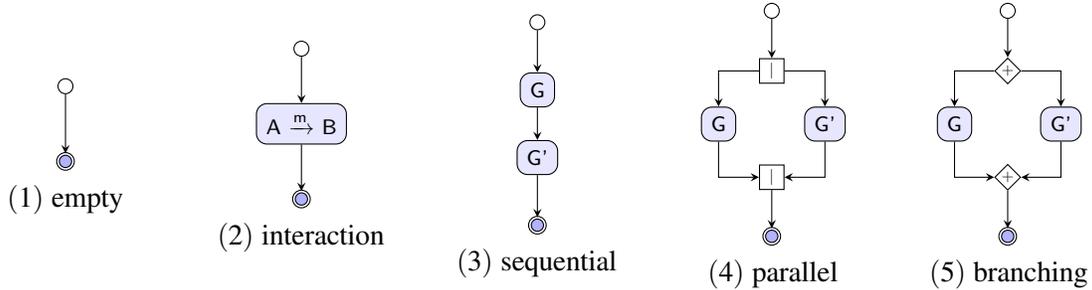

\begin{definition}[Global Choreography]\label{defGG}
\textit{A global choreography (g-choreography for short) is a term $\textsf{G}$ 
derived by the grammar
$$\textsf{G} ::= \ 0\  \mid \ A\xrightarrow{m} B \ \mid \  \textsf{G};\textsf{G'} \ \mid \  \textsf{G}| \textsf{G'} \ \mid \  \textsf{G}+\textsf{G'}$$
}
\end{definition}
%
The empty g-choreography is $0$; $A\xrightarrow{m} B$ is an interaction where message $m\in \mathcal{M}$ is sent from participant $A\in \mathcal{P}$ to participant $B\in \mathcal{P}$ ($A\neq B$); the operators $\_{;}\_$ and  $\_{\mid}\_$ and  $\_{+}\_$ allow us to compose g-choreographies sequentially, in parallel and in non-deterministic branches. 
The corresponding visual notation of g-choreographies is in Figure~\ref{VisualNotaionOfGG}. Circled and double circled nodes represent the unique initial and terminal node of each g-choreography. 

\paragraph{Pomsets for g-choreographies}
In~\cite{gt16} g-choreographies have been equipped with a pomset
semantics establishing causal dependencies among communication actions
on channels. Formally, the set of channels is
$\mathcal{C}=\{(A,B)\mid A,B\in \mathcal{P}, A\neq B\}$ (and we
abbreviate $(A,B)\in \mathcal{C}$ as $AB$). The set $\mathcal{L}$ of
labels
$$\mathcal{L}=\mathcal{L}^!\cup \mathcal{L}^? \ \ \textit{where} \ \ \mathcal{L}^!=\mathcal{C}\times \{!\}\times \mathcal{M} \ \ \textit{and}\ \ \mathcal{L}^?=\mathcal{C}\times \{?\}\times \mathcal{M}.$$
consists of the elements in the set $\mathcal{L}^!$ of outputs
representing send actions and $\mathcal{L}^?$ is the set of inputs
representing receive actions. We abbreviate
$(AB,!,m)\in \mathcal{L}^!$ and $(AB,?,m)\in \mathcal{L}^?$ with
$AB!m$ and $AB?m$ respectively.

\begin{definition}[Lposet \cite{gt16}]\label{defLposet}
  \textit{
    The subject \textit{sbj}(\_) and the object \textit{obj}(\_)of an action are  defined by
    \begin{align*}
      \textit{sbj}(AB!m)=A \ \ \textit{sbj}(AB?m)=B
      \qquad \text{and}\qquad
      &
      \textit{obj}(AB!m)=B \ \ \text{and}\ \ \textit{obj}(AB?m)=A
    \end{align*}
A labelled partially ordered set $r$ for g-choreographies (lposet) is a triple $(\mathcal{E}, \leq, \lambda)$, with $\mathcal{E}$ a set of events, $\leq \subseteq \mathcal{E}\times \mathcal{E}$ a partial order on $\mathcal{E}$, and $\lambda:\mathcal{E}\to \mathcal{L}$ a labelling function. }
\end{definition}

Note that Definition~\ref{defLposet} permits to have  $e\neq e'$, $\lambda(e)=\lambda(e')$, namely two events occur in different places with the same action. The relation $\leq$ is a partial order representing the causal dependencies among events. We use $e\to e'$ to denote $e\leq e'$. Also $\varepsilon$ denotes the empty lposet.

\begin{definition}[Isomorphism of lposets \cite{gt16}]\label{defIsoClassLposet}
\textit{
Two lposets $(\mathcal{E}, \leq, \lambda)$ and $(\mathcal{E'}, \leq', \lambda')$ are isomorphic iff there exists a bijection $\phi: \mathcal{E}\to \mathcal{E'}$ such that $e\leq e'$ iff $\phi(e)\leq' \phi(e')$ and $\lambda=\lambda' \circ \phi$.}
\end{definition}

\begin{definition}[Pomset \cite{gt16}]\label{defPomset}
\textit{
A partially-ordered multi-set $[\mathcal{E}, \leq, \lambda]$ (of actions), pomset for short, is the isomorphism class of an lposet $(\mathcal{E}, \leq, \lambda)$.}
\end{definition}
 
The advantage of using pomsets to formalise the semantics of g-choreographies is that partial orders  explicitly represent the causal dependencies among communications. 

Given a basic interaction $\textsf{G}=A\xrightarrow{m}B$, the pomset $\ldbrack \textsf{G} \rdbrack$ of $\textsf{G}$ is 
$$\ldbrack A\xrightarrow{m} B \rdbrack = \{[\mathcal{E}, \leq, \lambda]\}
=\{[(\{e_1,e_2\}, \{(e_1,e_1),(e_1,e_2),(e_2,e_2)\},\lambda)]\} \text{ where }
     \lambda =  \begin{cases}
                         e_1 \mapsto  AB!m\\
                         e_2 \mapsto  AB?m
                       \end{cases}
$$
Pomsets also have visual notation 
$$\ldbrack A\xrightarrow{m} B \rdbrack=\Bigg[ AB!m \rightarrow AB?m \Bigg]$$
where the pair of square brackets specifies the border of the pomset, the arrow from $AB!m$ to $AB?m$ shows the happen-before relationship between events. 

We now start to review the pomset semantics of  sequential and parallel composition. 
Given a natural number $n$, $\textbf{n}$ represents the singleton $\{n\}$ and $X \uplus Y$ the disjoint union of two sets $X$ and $Y$.
Also, given a function $f$ on $X$, we let $f\otimes \textbf{n}$ be the function extending $f$ to $X\times \textbf{n}$ mapping $(x,n)$ to $f(x)$; analogously, for a relation $R\subseteq X\times Y$, let $R\otimes \textbf{n}=\{((x,n),(y,n))\mid (x,y)\in R\}$ be the relation extending $R$ to $(X\times \textbf{n}) \times (Y\times \textbf{n})$.

\begin{definition}[Pomsets for sequential compositions]
\textit{Let $r=[\mathcal{E}, \leq, \lambda]$ and $r'=[\mathcal{E}', \leq', \lambda']$ be two pomsets.
For a pomset $r$ and a participant $A\in \mathcal{P}$, let $\mathcal{E}_{r,A}=\{e\in \mathcal{E}_r \mid \textsf{sbj}(\lambda_r(e))=A\}$ be the set of events of $A$ in $\mathcal{E}_r$. The semantics of the sequential composition $\textsf{seq}(r,r')$ of $r$ and $r'$ is defined as}
$$\textsf{seq}(r,r')=[\mathcal{E}\uplus \mathcal{E}', \leq_{\textsf{seq}}, (\lambda\otimes \textbf{1})\cup (\lambda' \otimes \textbf{2})],$$
\textit{where}
$$\leq_{\textsf{seq}}=\left( (\leq \otimes \textbf{1})\cup (\leq' \otimes \textbf{2})\cup \bigcup_{\textsf{A}\in \mathcal{P}}((\mathcal{E}_{r,\textsf{A}}\times \textbf{1})\times (\mathcal{E}_{r',\textsf{A}}\times \textbf{2})) \right)^\star$$
\textit{and $\star$ is the reflexive-transitive closure.}
\end{definition}

Now, we review parallel compositions of pomset. 


\begin{definition}[Pomsets for parallel compositions]
\textit{Let $r=[\mathcal{E}, \leq, \lambda]$ and $r'=[\mathcal{E}', \leq', \lambda']$ be two pomsets. The semantics of the parallel composition $\textsf{par}(r,r')$
 of $r$ and $r'$ is defined as}
$$\textsf{par}(r,r')=[\mathcal{E}\uplus \mathcal{E}', (\leq \otimes \textbf{1})\cup (\leq' \otimes \textbf{2}), (\lambda \otimes \textbf{1})\cup (\lambda' \otimes \textbf{2})].$$
\end{definition}

Roughly speaking, the sequential composition of pomsets $r$ and $r'$
adds causal dependencies between the communication events of $r$ and
$r'$ done by the same participant while parallel composition does not.

\paragraph{Well-formedness}

We relax the notion of well-formedness
of g-choreographies given in~\cite{gt16,gt17} by
considering only \emph{well-forkedness} and
\emph{well-branchedness}~\cite{gt16,gt17}:


\begin{definition}[well-forkedness]\label{genWf}
\textit{Pomsets $r=[\mathcal{E},\leq, \lambda]$ and $r'=[\mathcal{E'},\leq', \lambda']$ are well forked if }
$$\lambda(\mathcal{E})\cap \lambda'(\mathcal{E'}) \cap \mathcal{L}^?=\emptyset$$
\textit{we write $\textit{wf}(r,r')$ when $r$ and $r'$ are well-forked
  and, for} $\textsf{G}, \textsf{G'}, \textit{wf}(\textsf{G},\textsf{G'})$
\textit{when}
$\ldbrack \textsf{G} \rdbrack\neq \perp \land \ldbrack \textsf{G'}
\rdbrack\neq \perp \land \forall r\in \ldbrack \textsf{G} \rdbrack,
r'\in \ldbrack \textsf{G'} \rdbrack : \textit{wf}(r,r').$
\end{definition}

For a pomset $r=[\mathcal{E},\leq, \lambda]$, let
$\textsf{min}\ r=\{e\in\mathcal{E}\mid \nexists e'\in
\mathcal{E}:e'\neq e \land e'\leq e\}$ and, for a participant $A$, let
$r_{\downarrow A}=[\mathcal{E}_{r,A}, \leq \cap
(\mathcal{E}_{r,A},\mathcal{E}_{r,A}),\lambda\mid_{\mathcal{E}_{r,A}}]$
be the pomset projected on $A$ in $r$, where
$\mathcal{E}_{r,A}=\{e\in \mathcal{E}\mid \textit{sbj}(e)=A\}$,
$\lambda\mid_{\mathcal{E}_{r,A}}$ denotes the restriction of the
function $\lambda$ to the subset $\mathcal{E}_{r,A}$ of its domain.
%
The notion of well-branchedness requires the definition of
\emph{active} and \emph{passive} participants of branches of
g-choreographies.

\begin{definition}\it
  Given a branching g-choreography $\textsf{G}+\textsf{G'}$, let
  $\textsf{div}_A(\textsf{G},\textsf{G'})=(\tilde{l}_1,\tilde{l}_2)$
  with $\tilde{l}_1, \tilde{l}_2 \subseteq \mathcal{L}$
  defined as
  \[
    \tilde{l}_1=\bigcup_{r\in \ldbrack \textsf{G} \rdbrack} \lambda\mid_{\mathcal{E}_{r,A}}( \textsf{min}\ r_{\downarrow A})
    \qquad \text{and} \qquad
    \tilde{l}_2=\bigcup_{r'\in \ldbrack \textsf{G'} \rdbrack} \lambda\mid_{\mathcal{E}_{r',A}}( \textsf{min}\ r'_{\downarrow A})
  \]
  A participant $A \in \mathcal{P}$ is
  \begin{itemize}
  \item \emph{active in $\textsf{G}+\textsf{G'}$} if
    $\qquad \tilde{l}_1\cup \tilde{l}_2\subseteq \mathcal{L}^!
    \quad\land\quad
    \tilde{l}_1\cap \tilde{l}_2=\emptyset
    \quad\land\quad
    \tilde{l}_1\neq \emptyset
    \quad\land\quad
    \tilde{l}_2\neq \emptyset$
  \item \emph{passive in $\textsf{G}+\textsf{G'}$} if
    $\qquad\tilde{l}_1\cup \tilde{l}_2\subseteq \mathcal{L}^?
    \quad\land\quad
    \tilde{l}_1\cap \tilde{l}_2= \emptyset
    \quad\land\quad
    \tilde{l}_1\neq \emptyset
    \quad\land\quad
    \tilde{l}_2\neq \emptyset$.
  \end{itemize}
\end{definition}

An active (resp. passive) participant $A$ in $\textsf{G}+\textsf{G'}$
must send (resp. receive) different messages to (resp. from) other
participants at the branching starting points of g-choreographies. For
instance, $A$ is an active participant, while $B$ and $C$ are both
passive participants in the branching g-choreography $\textsf{G}$ in
Figure~\ref{fig:fakeErrSt}.

\begin{definition}[well-branchedness]\it
A g-choreography $\textsf{G'}+\textsf{G''}$ is well-branched if
\begin{enumerate}
\item there is at most one active participant in
  $\textsf{G}+\textsf{G'}$,
\item all the other participants of $\textsf{G}+\textsf{G'}$ are
  passive participants.
\end{enumerate}
We write $wb(\textsf{G},\textsf{G'})$ when $\textsf{G'}+\textsf{G''}$
is well-branched.
\end{definition}


We define a variant of semantics of g-choreographies based on considering only well-branchedness. 

\begin{definition}[Semantics of g-choreographies]\label{semanticOfGGSimple}
\textit{The semantics of a g-choreography is a family of pomsets defined as}
\begin{align*}
\ldbrack 0 \rdbrack &= \{\varepsilon\}\\
\ldbrack A\xrightarrow{m} B \rdbrack
&=\{[(\{e_1,e_2\}, \{(e_1,e_1),(e_1,e_2),(e_2,e_2)\},\lambda)]\} \textit{ where }
     \lambda =  \begin{cases}
                         e_1 \mapsto  AB!m\\
                         e_2 \mapsto  AB?m
                \end{cases}\\       
\ldbrack \textsf{G};\textsf{G'}  \rdbrack&= \{\textsf{seq}(r,r') \mid (r,r')\in \ldbrack \textsf{G}  \rdbrack \times \ldbrack \textsf{G'}  \rdbrack \}\\ 
\ldbrack \textsf{G} \mid \textsf{G'} \rdbrack&=
\begin{cases}
\{\textsf{par}(r,r') \mid (r,r')\in \ldbrack \textsf{G}  \rdbrack \times \ldbrack \textsf{G'}  \rdbrack \}   &\textit{if }  \textit{wf}(\textsf{G},\textsf{G'})\\
\perp  &\textit{otherwise}
\end{cases}\\
\ldbrack \textsf{G}+\textsf{G'} \rdbrack&= \begin{cases}
\ldbrack \textsf{G}  \rdbrack \cup \ldbrack \textsf{G'}  \rdbrack     &\textit{if } wb(\textsf{G},\textsf{G'})\ \\
\perp  &\textit{otherwise}
\end{cases}
\end{align*}
\end{definition}

\paragraph{Interface automata}

Interface automata \cite{de2001interface, de2005interface} play an important role in component-based design and verification. They capture input/output behaviours while the interfaces specify the possible interactions with the environment. 

\begin{definition}[Interface automata]
\textit{An interface automaton $M=(V, v_0,\mathcal{A},\mathcal{T})$ is a 4-tuple, where }

\begin{enumerate}
\item
\textit{$V$ is a finite set of states,}

\item
\textit{$v_0\in V$ is the initial state,}

\item
\textit{$\mathcal{A}=\mathcal{A}^I\cup \mathcal{A}^O\cup \mathcal{A}^H$ is the set of actions, 
where $\mathcal{A}^I$, $\mathcal{A}^O$ and $\mathcal{A}^H$ are pair-wise disjoint sets of input, output and internal actions, respectively,}

\item
\textit{$\mathcal{T}\subseteq V\times \mathcal{A}\times  V$ is a set of transitions.}
\end{enumerate}
\end{definition}

\noindent
We write $v\xrightarrow{a}v'$ instead of $(v,a,v')\in \mathcal{T}$ when the set of transitions $\mathcal{T}$ is clear from the context.
Interface automata are able to synchronously interact with each other according to Definition~\ref{comp}. 
\begin{definition}[Composability and Product]\label{comp}
\textit{Two interface automata $M$ and $N$ are composable if 
$$\mathcal{A}_M^H \cap \mathcal{A}_N= \emptyset,  \quad \mathcal{A}_M^I \cap \mathcal{A}_N^I=\emptyset,\quad
\mathcal{A}_M^O \cap \mathcal{A}_N^O= \emptyset,  \quad \mathcal{A}_N^H \cap \mathcal{A}_M=\emptyset.$$
We define $\textit{
shared}(M,N)=\mathcal{A}_M\cap \mathcal{A}_N$. 
If two interface automata $M$ and $N$ are composable, then we have $\textit{shared}(M,N)=(\mathcal{A}_M^I\cap \mathcal{A}_N^O)\cup (\mathcal{A}_M^O\cap \mathcal{A}_N^I)$. If two interface automata $M$ and $N$ are composable, their synchronous product $M\otimes N$ is the interface automaton $M\otimes N=(V, v_0, \mathcal{A}, \mathcal{T})$,}
\begin{align*}
V&=V_M \times V_N,\\
{v_0}&=({v_0}_M, {v_0}_N),\\
\mathcal{A}&=\mathcal{A}^I\cup \mathcal{A}^O \cup \mathcal{A}^H,\\
\mathcal{A}^I&=(\mathcal{A}_M^I\cup \mathcal{A}_N^I)\setminus \textit{shared(M,N)},\\
\mathcal{A}^O&=(\mathcal{A}_M^O\cup \mathcal{A}_N^O)\setminus \textit{shared(M,N)},\\
\mathcal{A}^H&=\mathcal{A}_M^H\cup \mathcal{A}_N^H\cup \textit{shared(M,N)},\\
\mathcal{T}&=\{(v,u)\xrightarrow{a}(v',u) \mid (v\xrightarrow{a} v')\in \mathcal{T}_M \land a \notin \textit{shared}(M,N)\land u\in V_N\}\\
&\cup\{(v,u)\xrightarrow{a}(v,u') \mid (u\xrightarrow{a} u')\in \mathcal{T}_N \land a \notin \textit{shared}(M,N)\land v\in V_M\}\\
&\cup\{(v,u)\xrightarrow{a}(v',u') \mid (v\xrightarrow{a} v')\in \mathcal{T}_M \land (u\xrightarrow{a} u')\in \mathcal{T}_N \land a\in \textit{shared}(M,N)\}
\end{align*}
\end{definition}

\noindent
Interactions of interface automata may lead to error states, which correspond to potential deadlocks.

\begin{definition}[Error states]\label{defErrorState0}
\textit{The set of error states of the product of two interface automata $M$ and $N$ is defined by}
\begin{align*}
\textit{Error}(M,N) & =
                              \big\{(v,u)\in V_M \times V_N \mid \\
                    &(\exists a \in \textit{shared}(M,N)\land a \in \mathcal{A}_M^O: (v\xrightarrow{a}v')\in \mathcal{T}_M \land a\in \mathcal{A}_N^I: (u\xrightarrow{a}u')\notin \mathcal{T}_N)\lor
    \\
                        & (\exists a \in \textit{shared}(M,N)\land a \in \mathcal{A}_N^O: (u\xrightarrow{a}u')\in \mathcal{T}_N \land a \in\mathcal{A}_M^I:(v\xrightarrow{a}v')\notin \mathcal{T}_M)\big\}
\end{align*}

\end{definition}

\newcommand{\trproj}[3]{\textsf{proj}_{\achan[#2][#3]}(\ifempy{#1}{\_}{#1})}

\section{Group Interface Automata}\label{sectGIA}

We propose an extension 
of interface automata called group interface automata (GIA) as a convenient representation of local views of choreographies.

\begin{definition}[Group interface automata]
\textit{A group interface automaton $\mathcal{I}=(V, v_0,\mathcal{G},\mathcal{A},\mathcal{T})$ is a 5-tuple defined as follows, }

\begin{enumerate}
\item \textit{$V$ is a finite set of states,}

\item
\textit{$v_0\in V$ is the initial state,}

\item
\textit{$\mathcal{G}\subseteq \mathcal{P}$ is a finite set of participants,}

\item
\textit{$\mathcal{A}=\mathcal{A}^I\cup \mathcal{A}^O\cup \mathcal{A}^H$ is a set of interfaces (actions), where 
$$\mathcal{A}^I=(\mathcal{P}\setminus\mathcal{G})\times \mathcal{G}\times \{?\}\times \mathcal{M},\quad \mathcal{A}^O=\mathcal{G}\times (\mathcal{P}\setminus\mathcal{G})\times \{!\}\times \mathcal{M},\quad \mathcal{A}^H=\mathcal{G}\times \mathcal{G}\times \{!?\} \times \mathcal{M},$$}
\item
\textit{$\mathcal{T}\subseteq V\times \mathcal{A}\cup\{\tau\}\times  V$ is a set of transitions.}
\end{enumerate}
\end{definition}

\noindent 
The symbols $``?", ``!", ``!?"$ mark input, output and internal actions respectively. Action $\tau$ represents a special internal computation. Given a GIA $\mathcal{I}$, we write $\mathcal{I}_{\mathcal{G}}$ when we want to highlight the group of participants in $\mathcal{I}$  
and also $\mathcal{I}_A$ instead of $\mathcal{I}_{\{A\}}$. 
 $AB?m$ and $AB!m$ and $AB!?m$ abbreviate $(A,B,?,m)\in\mathcal{A}^I$, $(A,B,!,m)\in\mathcal{A}^O$, $(A,B,!?,m)\in\mathcal{A}^H$, respectively. The subject and object of transition actions are
$$\textit{sbj}(AB!m)=\textit{obj}(AB?m)=\textit{sbj}(AB!?m)= A,  \qquad \textit{obj}(AB!m)=\textit{sbj}(AB?m)=\textit{obj}(AB!?m) = B$$

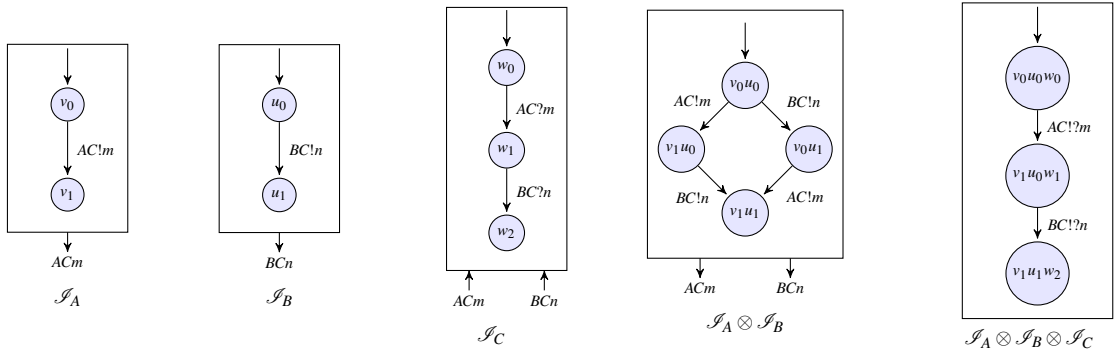
\begin{figure}[htbp]
\begin{minipage}{0.15\textwidth}
\begin{tikzpicture}[shorten >=1pt,node distance=1.2cm,>=stealth',auto,
                                    every state/.style={thin,fill=blue!10}]      
\draw (-0.8,-1.7) rectangle (0.8,0.8);
\draw  [->] (0.0,-1.7)--(0.0,-2.0);
\node at (0.0,-2.1) {\tiny $ACm$};                                                             
\node[state,initial above,inner sep=2pt,minimum size=0pt]        (q_0)                           {\tiny $v_0$};  
\node[state,inner sep=2pt,minimum size=0pt]  (q_1)  [below of=q_0]     {\tiny $v_1$}; 
\path[->]       
             (q_0)    edge    node{\tiny $AC!m$}     (q_1);
\end{tikzpicture}
\centering
\\ \scriptsize $\mathcal{I}_A$
\end{minipage}
\begin{minipage}{0.19\textwidth}
\begin{tikzpicture}[shorten >=1pt,node distance=1.2cm,>=stealth',auto,
                                    every state/.style={thin,fill=blue!10}]      
\draw (-0.8,-1.7) rectangle (0.8,0.8);
\draw  [->] (0.0,-1.7)--(0.0,-2.0);
\node at (0.0,-2.1) {\tiny $BCn$};                                                             
\node[state,initial above,inner sep=2pt,minimum size=0pt]        (q_0)                           {\tiny $u_0$};  
\node[state,inner sep=2pt,minimum size=0pt]  (q_1)  [below of=q_0]     {\tiny $u_1$}; 
\path[->]       
             (q_0)    edge    node{\tiny $BC!n$}     (q_1);
\end{tikzpicture}
\centering
\\ \scriptsize $\mathcal{I}_B$
\end{minipage}
\begin{minipage}{0.15\textwidth}
\begin{tikzpicture}[shorten >=1pt,node distance=1.1cm,>=stealth',auto,
                                    every state/.style={thin,fill=blue!10}]     
\node at (-1.1,-0.8) {}; 
\draw (-0.8,-2.7) rectangle (0.8,0.8);  
\draw  [<-] (-0.5,-2.7)--(-0.5,-3.0);
\draw  [<-] (0.5,-2.7)--(0.5,-3.0);
\node at (-0.5,-3.1) {\tiny $ACm$};              
\node at (0.5,-3.1) {\tiny $BCn$};   

\node[state,initial above,inner sep=2pt,minimum size=0pt]        (q_0)                           {\tiny $w_0$};  
\node[state,inner sep=2pt,minimum size=0pt]  (q_1)  [below  of=q_0]     {\tiny $w_1$}; 
\node[state,inner sep=2pt,minimum size=0pt]  (q_2)  [below  of=q_1]     {\tiny $w_2$}; 

\path[->]       
             (q_0)     edge   node{\tiny ${AC?m}$}     (q_1)  
             (q_1)     edge   node{\tiny ${BC?n}$}     (q_2);
\end{tikzpicture}
\centering
\\ \scriptsize $\mathcal{I}_C$
\end{minipage}
\begin{minipage}{0.26\textwidth}
\begin{tikzpicture}[shorten >=1pt,node distance=1.2cm,>=stealth',auto,
                                    every state/.style={thin,fill=blue!10},
box/.style={rectangle, draw=red!60, fill=red!5, very thick, minimum size=7mm}]     
\node at (-1.1,-0.8) {}; 
\draw (-1.3,-2.3) rectangle (1.3,1.0);  
\draw  [->] (-0.6,-2.3)--(-0.6,-2.6);
\draw  [->] (0.6,-2.3)--(0.6,-2.6);
\node at (-0.6,-2.7) {\tiny $ACm$};  
\node at (0.6,-2.7) {\tiny $BCn$};  

\node at (-0.7,-0.2) {\tiny $AC!m$};            
\node at (-0.7,-1.5) {\tiny $BC!n$};     

\node[state,initial above,inner sep=2pt,minimum size=0pt]        (q_0)                           {\tiny $v_0u_0$};  
\node[state,inner sep=2pt,minimum size=0pt]  (q_1)  [below left of=q_0]     {\tiny $v_1u_0$}; 
\node[state,inner sep=2pt,minimum size=0pt]  (q_2)  [below right of=q_0]     {\tiny $v_0u_1$}; 
\node[state,inner sep=2pt,minimum size=0pt]  (q_3)  [below right of=q_1]     {\tiny $v_1u_1$}; 

\path[->]       
             (q_0)     edge   node{}     (q_1)
                       edge   node{\tiny ${BC!n}$}     (q_2)
             (q_1)     edge   node{}     (q_3)
             (q_2)     edge   node{\tiny ${AC!m}$}     (q_3);
\end{tikzpicture}
\centering
\\ \scriptsize $\mathcal{I}_A\otimes \mathcal{I}_B$
\end{minipage}
\begin{minipage}{0.20\textwidth}
\begin{tikzpicture}[shorten >=1pt,node distance=1.3cm,>=stealth',auto,
                                    every state/.style={thin,fill=blue!10},
box/.style={rectangle, draw=red!60, fill=red!5, very thick, minimum size=7mm}]     
\node at (-1.1,-0.8) {}; 
\draw (-1.0,-3.2) rectangle (1.0,1.0);  
\node[state,initial above,inner sep=2pt,minimum size=0pt]        (q_0)                           {\tiny $v_0u_0w_0$};  
\node[state,inner sep=2pt,minimum size=0pt]  (q_1)  [below  of=q_0]     {\tiny $v_1u_0w_1$}; 
\node[state,inner sep=2pt,minimum size=0pt]  (q_2)  [below  of=q_1]     {\tiny $v_1u_1w_2$}; 

\path[->]       
             (q_0)     edge   node{\tiny ${AC!?m}$}     (q_1)  
             (q_1)     edge   node{\tiny ${BC!?n}$}     (q_2);
\end{tikzpicture}
\centering
\\ \scriptsize $\mathcal{I}_A\otimes \mathcal{I}_B\otimes \mathcal{I}_C$
\end{minipage}
\caption{The $\otimes$-product among three composable GIA}\label{fig:productGIA}
\end{figure}

\noindent
Figure~\ref{fig:productGIA} shows some instances of GIA. Let us comment on $\mathcal{I}_A$ and $\mathcal{I}_C$. The rectangles specify the scope of the machines, the arrow labeled $ACm$ leaving from the border of $\mathcal{I}_A$ and the arrow labeled $ACm$ leading to the border of $\mathcal{I}_C$ represent the output and input interfaces of $\mathcal{I}_A$ and $\mathcal{I}_C$  respectively.



\begin{definition}[Shared interface]
\textit{Given two GIA $\mathcal{I}=(V,v_0, \mathcal{G},\mathcal{A}, \mathcal{T})$ and $\mathcal{I'}=(V',v_0', \mathcal{G'},\mathcal{A'},\mathcal{T'})$,  the shared inputs, shared outputs and shared internals between $\mathcal{I}$ and $\mathcal{I'}$ are $\textit{si}(\mathcal{I},\mathcal{I'})$, $\textit{so}(\mathcal{I},\mathcal{I'})$, $\textit{sh}(\mathcal{I},\mathcal{I'})$ respectively, defined as}
\begin{align*}
\textit{si}(\mathcal{I},\mathcal{I'})&=\{AB?m\in \mathcal{A}\mid AB!m\in \mathcal{A'}\}\cup \{AB?m\in \mathcal{A'}\mid AB!m\in \mathcal{A}\},\\
\textit{so}(\mathcal{I},\mathcal{I'})&=\{AB!m\in \mathcal{A}\mid AB?m\in \mathcal{A'}\}\cup \{AB!m\in \mathcal{A'}\mid AB?m\in \mathcal{A}\},\\
\textit{sh}(\mathcal{I},\mathcal{I'})&=\{AB!?m \mid AB!m\in \mathcal{A}\land AB?m\in \mathcal{A'} \}\cup \{AB!?m\mid AB!m\in \mathcal{A'}\land AB?m\in \mathcal{A} \}.
\end{align*}
\end{definition}

The composition of GIA assumes that each pair of automata $\mathcal{I}$ and $\mathcal{I'}$ uses a channel from $\mathcal{I}$ to $\mathcal{I'}$ (resp. $\mathcal{I'}$ to $\mathcal{I}$) to send messages from $\mathcal{I}$ to $\mathcal{I'}$ (resp. $\mathcal{I'}$ to $\mathcal{I}$), In CFSMs~\cite{brand1983communicating} asynchrony is realised by means of buffered channels, namely channels that allow messages from the sender to be stored and consumed later by the receiver. For our purposes, it is enough to assume that the size of the buffers of each channel is one. As describe below, our definition of composition induces one-size buffer channels between each two participants, When an automata wants to execute on output, the message is dispatched to the buffer connected between sender and receiver provided the buffer is empty. Dually, a machine who wants to receive a message access its buffer first, then consumes the message if there is any.

\begin{definition}[$\otimes$-product]\label{productOfAIA}
\textit{Two GIA $\mathcal{I'}=(V',v_0', \mathcal{G'},\mathcal{A'}, \mathcal{T'})$ and $\mathcal{I''}=(V'',v_0'', \mathcal{G''},\mathcal{A''},\mathcal{T''})$ are composable if $\mathcal{G'}\cap \mathcal{G''}=\emptyset$.}
\textit{If two GIA $\mathcal{I'}$ and $\mathcal{I''}$ are composable, $\mathcal{I'}\otimes \mathcal{I''}$ is the GIA $\mathcal{I}=(V' \times V'',(v_0', v_0''),\mathcal{G'} \cup \mathcal{G''},\mathcal{A}^I \cup \mathcal{A}^O  \cup \mathcal{A}^H, \mathcal{T})$, where}
\begin{align*}
\mathcal{A}^I&=(\mathcal{A'}^I \cup \mathcal{A''}^I) \setminus \textit{si}(\mathcal{I'},\mathcal{I''}),\\
\mathcal{A}^O&=(\mathcal{A'}^O \cup \mathcal{A''}^O) \setminus \textit{so}(\mathcal{I'},\mathcal{I''}),\\
\mathcal{A}^H&=\mathcal{A'}^H \cup \mathcal{A''}^H \cup\textit{sh}(\mathcal{I'},\mathcal{I''}),\\
\mathcal{T}
&=\{(v',v'')\xrightarrow{\alpha}(u',v'') \mid (v'\xrightarrow{\alpha} u')\in \mathcal{T'} \land \alpha \notin (\textit{si}(\mathcal{I'},\mathcal{I''}) \cup \textit{so}(\mathcal{I'},\mathcal{I''})) \land v''\in V''\}\\
&\cup\{(v',v'')\xrightarrow{\alpha}(v',u'') \mid  (v''\xrightarrow{\alpha} u'')\in \mathcal{T''} \land \alpha \notin (\textit{si}(\mathcal{I'},\mathcal{I''}) \cup \textit{so}(\mathcal{I'},\mathcal{I''})) \land v'\in V'\}\\
&\cup\{(v',v'')\xrightarrow{AB!?m}(u',u'')\mid (v'\xrightarrow{AB!m}u')\in \mathcal{T'} \land (v''\xrightarrow{AB?m}u'')\in \mathcal{T''} \}\\
&\cup\{(v',v'')\xrightarrow{AB!?m}(u',u'')\mid (v''\xrightarrow{AB!m}u'')\in \mathcal{T''} \land (v'\xrightarrow{AB?m}u')\in \mathcal{T'} \}.
\end{align*}
\end{definition}

Interactions among GIA are captured by their product. Figure~\ref{fig:productGIA}  shows the $\otimes$-products among three pair-wise composable GIA.
%
The product $\mathcal{I}_A \otimes \mathcal{I}_B$ yields the
interaction between GIA $\mathcal{I}_A$ and
$\mathcal{I}_B$. 
Due to the sets of interfaces of $\mathcal{I}_A$ and $\mathcal{I}_B$
being disjoint, then, according to Definition~\ref{productOfAIA}, the
structure of $\mathcal{I}_A \otimes \mathcal{I}_B$ is interleaving the
transitions of $\mathcal{I}_A$ and $\mathcal{I}_B$.
$\mathcal{I}_A \otimes \mathcal{I}_B \otimes \mathcal{I}_C$ has only
internal interfaces since the set of interfaces of
$\mathcal{I}_A \otimes \mathcal{I}_B$ and $\mathcal{I}_C$ are
complementary. Moreover, the operator $\_\otimes\_$ is commutative and
associative, which can be proved by showing the bisimulation of sets of states between two GIA. 
Notice that the composition of interface automata is \quo{optimistic}
in the sense that it allows one to make automata interact as long as
they can execute some traces avoiding error states. This does not fit
choreographic approaches like ours. In fact, the idea of choregraphies
is to specify systems that never run into error states. Hence,
composition in GIA tries to single out all the error states emerging
from interactions.

\newcommand{\atrace}[1][t]{\mathsf{#1}}
\newcommand{\actoccur}[2]{\#(\atrace[{#1}],#2)}

\begin{theorem}[Commutativity and associativity]
Let $\mathcal{I}$, $\mathcal{I'}$ and $\mathcal{I''}$ be three GIA, then $\mathcal{I}\otimes \mathcal{I'} =\mathcal{I'}\otimes \mathcal{I}$. If $\mathcal{I}$, $\mathcal{I'}$ and $\mathcal{I''}$ are pair-wise composable, then $(\mathcal{I}\otimes \mathcal{I''})\otimes \mathcal{I'} = \mathcal{I}\otimes (\mathcal{I'}\otimes \mathcal{I''}).$
\end{theorem}

Next, similarly to error states in interface automata, we define error states in the $\otimes$-product of GIA. 
Let the dual of an action be defined as
$$\textit{dual}(AB!m)=AB?m\qquad \textit{dual}(AB?m)=AB!m\qquad \textit{dual}(AB!?m)=AB!?m$$
and the dual of a string $\alpha_0\alpha_1\cdots\alpha_n$ as $\textit{dual}(\alpha_0\alpha_1\cdots\alpha_n)=\textit{dual}(\alpha_0)\textit{dual}(\alpha_1)\cdots\textit{dual}(\alpha_n). $
%
Given a transition $v\xrightarrow{\alpha}u$, we define 
\begin{equation}
\#(v\xrightarrow{\alpha}u, A, B)=
\begin{cases}
\alpha &\textit{if }  \textit{sbj}(\alpha)=A \textit{ and } \textit{obj}(\alpha)=B \\
\tau & \textit{otherwise} 
\end{cases}
\end{equation}
where we overload $\tau$ to represent the empty string.
Then, we extend $\#(\_,\_,\_)$ to sequences of transitions
$\textsf{t} = v \xrightarrow{\alpha_0} \xrightarrow{\alpha_1} \cdots
\xrightarrow{\alpha_n}u$ as follows
$$\#(\textsf{t}, A, B)=\#(\alpha_0,A,B)\#(\alpha_i,A,B)\cdots\#(\alpha_n,A,B).$$
Given a string $\omega$, the prefixes of $\omega$ are $\textsf{Pref}(\omega)=\{\mu | \exists \nu\,.\,  \omega=\mu\nu\}$.
Given a non-empty  string $\omega\alpha$, we define an operation that removes the last action if it is an output, where
\begin{equation}
\textsf{ro}(\omega\alpha)=
\begin{cases}
\omega &\textit{if }  \alpha \in \mathcal{L}^! \\ 
\omega\alpha & \textit{otherwise} 
\end{cases}
\end{equation}

\begin{definition}[Error states of the $\otimes$-product]\label{defErrorState}\it
  Let $v$ and $v'$ be states respectively of GIAs $\mathcal{I}$
  and $\mathcal{I'}$.
  We say that $v$ has an \emph{unmatched shared output by $v'$} if
  there exists $AB!m \in \textit{so} (\mathcal{I}, \mathcal{I'})$ and
  a sequence of transitions
  $\atrace = v \xrightarrow{AB!m}
  \xrightarrow{\beta_0}\cdots\xrightarrow{\beta_k} u \in \mathcal{I}$
  then for all
  $\atrace[t'] =
  v' \xrightarrow{\alpha_0}\cdots\xrightarrow{\alpha_n}
  u' \in \mathcal{I'}$ either of the following conditions holds
  \begin{enumerate}
  \item\label{def:1}
    $\alpha_i \neq AB?m$ for all $0 \leq i \leq n$
  \item \label{def:2} there is $0 < i \leq n$ such that
    $\alpha_i = AB?m$; in this case, let $\hat h$ be the minimal index
    such that $\alpha_{\hat h} = AB?m$, then there is
    $0 \leq i < \hat h$ for which
    $\alpha_i \in \textit{si}(\mathcal{I},\mathcal{I'})\cup
    \textit{so}(\mathcal{I},\mathcal{I'})$ with
    $\textit{sbj}(\alpha_i) = B$ and
    \[
      \textit{obj}(\alpha_i) =  C \neq A \implies
      \textsf{ro}(\#(\atrace[t']_{\leq \hat h}, B, C)) \not\in \textsf{Pref}(\textit{dual}(\#(\atrace[t], C, B)))
    \]
  \end{enumerate}
  A state $(v,v')$ of the $\otimes$-product of two GIA $\mathcal{I}$
  and $\mathcal{I'}$ is an \emph{error state of
    $\mathcal{I}\otimes \mathcal{I'}$} if $v$ has unmatched shared
  output by $v'$ or $v'$ has an unmatched shared outputs by $v$.
  We denote the set of error states of
  $\mathcal{I}\otimes \mathcal{I'}$ with
  $\textit{Error}(\mathcal{I},\mathcal{I'})$.
\end{definition}

\begin{figure}[h]
\begin{minipage}{0.19\textwidth}
\begin{tikzpicture}[shorten >=1pt,node distance=1.0cm,>=stealth',auto,
                                    every state/.style={thin,fill=blue!10}]      
\draw (-1.0,-2.3) rectangle (1.0,0.8);
\draw  [->] (-0.6,-2.3)--(-0.6,-2.6);
\draw  [->] (0.6,-2.3)--(0.6,-2.6);
\node at (-0.6,-2.7) {\tiny $ABm$};                                                             
\node at (0.6,-2.7) {\tiny $ABn$};                                                             
\node[state,initial above,inner sep=2pt,minimum size=0pt]        (q_0)                           {\tiny $v_0$};  
\node[state,inner sep=2pt,minimum size=0pt]  (q_1)  [below  of=q_0]     {\tiny $v_1$}; 
\node[state,inner sep=2pt,minimum size=0pt]  (q_2)  [below  of=q_1]     {\tiny $v_2$};  

\path[->]       
             (q_0)   edge  node{\tiny $AB!m$}     (q_1)           		
             (q_1)   edge  node{\tiny $AB!n$}     (q_2);
\end{tikzpicture}
\centering
\\ \scriptsize $\mathcal{I}_{A}$
\end{minipage}
\begin{minipage}{0.19\textwidth}
\begin{tikzpicture}[shorten >=1pt,node distance=1.1cm,>=stealth',auto,
                                    every state/.style={thin,fill=blue!10}]      
\draw (-1.0,-1.7) rectangle (1.0,0.9);
\draw  [<-] (-0.6,-1.7)--(-0.6,-2.0);
\draw  [<-] (0.6,-1.7)--(0.6,-2.0);
\node at (-0.6,-2.1) {\tiny $ABm$};  
\node at (0.6,-2.1) {\tiny $ABn$};                                                         \node at (-0.6,-0.6) {\tiny $AB?m$};                                                             
\node[state,initial above,inner sep=2pt,minimum size=0pt]        (q_0)                           {\tiny $u_0$};  
\node[state,inner sep=2pt,minimum size=0pt]  (q_1)  [below   of=q_0]     {\tiny $u_1$}; 

\path[->]       
             (q_0)   edge[bend right]    node{}     (q_1)   
                     edge[bend left]    node{\tiny $AB?n$}     (q_1);
\end{tikzpicture}
\centering
\\ \scriptsize $\mathcal{I}_{B}$
\end{minipage}
\begin{minipage}{0.19\textwidth}
\begin{tikzpicture}[shorten >=1pt,node distance=1.0cm,>=stealth',auto,
                                    every state/.style={thin,fill=blue!10}]      
\draw (-1.0,-2.3) rectangle (1.0,0.8);
\draw  [<-] (-0.6,-2.3)--(-0.6,-2.6);
\draw  [<-] (0.6,-2.3)--(0.6,-2.6);
\node at (-0.6,-2.7) {\tiny $ABm$};                                                             
\node at (0.6,-2.7) {\tiny $ABn$};                                                             
\node[state,initial above,inner sep=2pt,minimum size=0pt]        (q_0)                           {\tiny $u'_0$};  
\node[state,inner sep=2pt,minimum size=0pt]  (q_1)  [below   of=q_0]     {\tiny $u'_1$}; 
\node[state,inner sep=2pt,minimum size=0pt]  (q_2)  [below   of=q_1]     {\tiny $u'_2$};

\path[->]       
             (q_0)     edge    node{\tiny $AB?n$}     (q_1)
             (q_1)     edge    node{\tiny $AB?m$}     (q_2);
\end{tikzpicture}
\centering
\\ \scriptsize $\mathcal{I'}_{B}$
\end{minipage}
\begin{minipage}{0.19\textwidth}
\begin{tikzpicture}[shorten >=1pt,node distance=1.3cm,>=stealth',auto,
                                    every state/.style={thin,fill=blue!10},
box/.style={rectangle, draw=red!60, fill=red!5, very thick, minimum size=7mm}]      
\draw (-1.0,-1.8) rectangle (1.0,0.9);
\node[state,initial above,inner sep=2pt,minimum size=0pt]        (q_0)                           {\tiny $v_0u_0$};  
\node[box,inner sep=2pt,minimum size=0pt]  (q_1)  [below  of=q_0]     {\tiny $v_1u_1$}; 

\path[->]       
             (q_0)   edge  node{\tiny $AB!?m$}     (q_1)  ;  
\end{tikzpicture}
\centering
\\ \scriptsize $\mathcal{I}_{A}\otimes \mathcal{I}_{B}$
\end{minipage}
\begin{minipage}{0.19\textwidth}
\begin{tikzpicture}[shorten >=1pt,node distance=1.3cm,>=stealth',auto,
                                    every state/.style={thin,fill=blue!10},
box/.style={rectangle, draw=red!60, fill=red!5, very thick, minimum size=7mm}]      
\draw (-1.0,-1.0) rectangle (1.0,0.9);
\node[box,initial above,inner sep=2pt,minimum size=0pt]        (q_0)                           {\tiny $v_0u'_0$};  
\end{tikzpicture}
\centering
\\ \scriptsize $\mathcal{I}_{A}\otimes \mathcal{I'}_B$
\end{minipage}\newline

\begin{minipage}{0.33\textwidth}
\begin{tikzpicture}[shorten >=1pt,node distance=1.0cm,>=stealth',auto,
                                    every state/.style={thin,fill=blue!10}]      
\draw (-1.0,-3.3) rectangle (1.0,0.8);
\draw  [->] (-0.6,-3.3)--(-0.6,-3.6);
\draw  [->] (0.6,-3.3)--(0.6,-3.6);
\draw  [->] (0.0,-3.3)--(0.0,-3.6);
\node at (-0.6,-3.7) {\tiny $ACm$};               
\node at (-0.0,-3.7) {\tiny $ACx$};              
\node at (0.6,-3.7) {\tiny $BCy$};                                                             
\node at (-0.6,-2.6) {\tiny $AC!m$};                                                             
\node[state,initial above,inner sep=2pt,minimum size=0pt]        (q_0)                           {\tiny $v_0$};  
\node[state,inner sep=2pt,minimum size=0pt]  (q_1)  [below  of=q_0]     {\tiny $v_1$}; 
\node[state,inner sep=2pt,minimum size=0pt]  (q_2)  [below  of=q_1]     {\tiny $v_2$};  
\node[state,inner sep=2pt,minimum size=0pt]  (q_3)  [below  of=q_2]     {\tiny $v_3$};  

\path[->]       
             (q_0)   edge  node{\tiny $AC!m$}     (q_1)           		
             (q_1)   edge  node{\tiny $BC!y$}     (q_2)
             (q_2)   edge [bend right]  node{}     (q_3)
             (q_2)   edge [bend left]  node{\tiny $AC!x$}     (q_3);
\end{tikzpicture}
\centering
\\ \scriptsize $\mathcal{I}_{\{A,B\}}$
\end{minipage}
\begin{minipage}{0.33\textwidth}
\begin{tikzpicture}[shorten >=1pt,node distance=1.0cm,>=stealth',auto,
                                    every state/.style={thin,fill=blue!10}]      
\draw (-1.0,-3.3) rectangle (1.0,0.8);
\draw  [<-] (-0.6,-3.3)--(-0.6,-3.6);
\draw  [<-] (0.6,-3.3)--(0.6,-3.6);
\draw  [<-] (0.0,-3.3)--(0.0,-3.6);
\node at (-0.6,-3.7) {\tiny $ACm$};              
\node at (-0.0,-3.7) {\tiny $ACx$};              
\node at (0.6,-3.7) {\tiny $BCy$};                                                             
\node[state,initial above,inner sep=2pt,minimum size=0pt]        (q_0)                           {\tiny $u_0$};  
\node[state,inner sep=2pt,minimum size=0pt]  (q_1)  [below  of=q_0]     {\tiny $u_1$}; 
\node[state,inner sep=2pt,minimum size=0pt]  (q_2)  [below  of=q_1]     {\tiny $u_2$};  
\node[state,inner sep=2pt,minimum size=0pt]  (q_3)  [below  of=q_2]     {\tiny $u_3$};  

\path[->]       
             (q_0)   edge  node{\tiny $AC?m$}     (q_1)           		
             (q_1)   edge  node{\tiny $AC?x$}     (q_2)
             (q_2)   edge  node{\tiny $BC?y$}     (q_3);
\end{tikzpicture}
\centering
\\ \scriptsize $\mathcal{I}_{C}$
\end{minipage}
\begin{minipage}{0.33\textwidth}
\begin{tikzpicture}[shorten >=1pt,node distance=1.3cm,>=stealth',auto,
                                    every state/.style={thin,fill=blue!10},
box/.style={rectangle, draw=red!60, fill=red!5, very thick, minimum size=7mm}]      
\draw (-1.0,-1.8) rectangle (1.0,0.9);
\node[state,initial above,inner sep=2pt,minimum size=0pt]        (q_0)                           {\tiny $v_0u_0$};  
\node[box,inner sep=2pt,minimum size=0pt]  (q_1)  [below  of=q_0]     {\tiny $v_1u_1$}; 

\path[->]       
             (q_0)   edge  node{\tiny $AC!?m$}     (q_1)  ;       
\end{tikzpicture}
\centering
\\ \scriptsize $\mathcal{I}_{\{A,B\}}\otimes \mathcal{I}_{C}$
\end{minipage}
\caption{Error states in GIA}\label{ex:defErrorStates}
\end{figure}
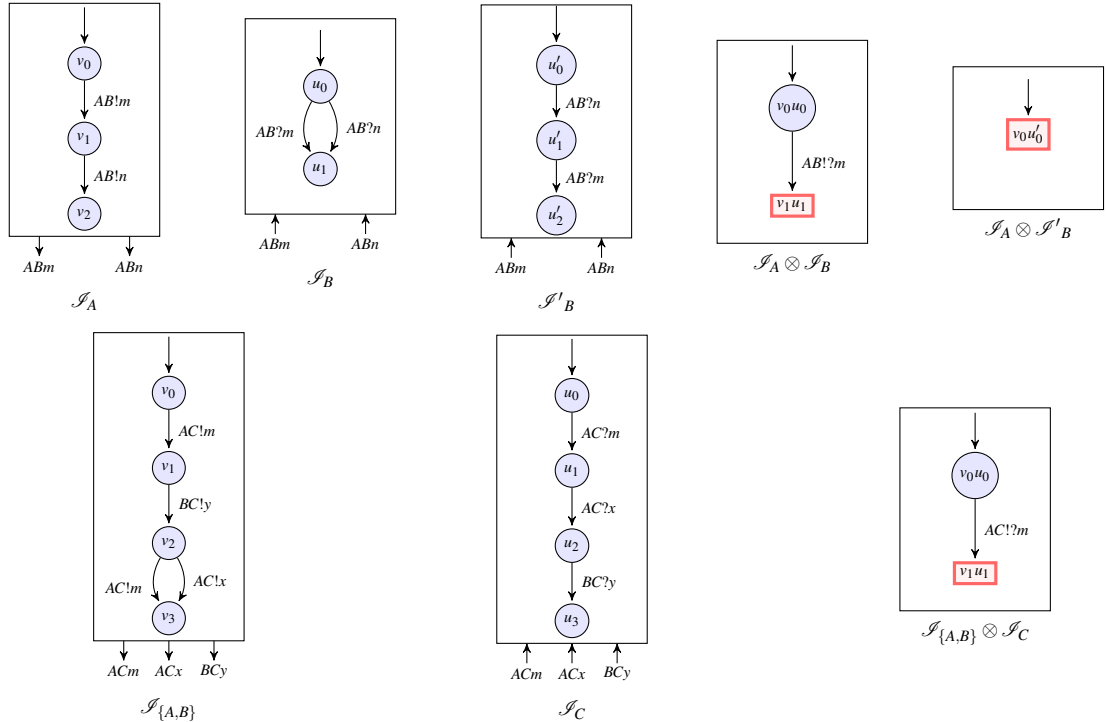

The definition of error states detects the situation where at least one shared output is not consumed. Figure~\ref{ex:defErrorStates} shows three classes of error states represented by red boxes.
$(1)$ The state $v_1u_1$ of $\mathcal{I}_A\otimes \mathcal{I}_B$ is an error state (by \eqref{def:1} in Definition~\ref{defErrorState}) since the shared output $AB!n$ in state $v_1$ does not have a corresponding input $AB?m$ in state $u_1$. In other words, when $\mathcal{I}_A$ wants to send $n$ to $\mathcal{I}_B$ at state $v_1$, $\mathcal{I}_B$ will never receive $n$ at state $u_1$.    
$(2)$ The initial state $v_0u'_0$ of $\mathcal{I}_A\otimes \mathcal{I'}_B$ is an error state (by \eqref{def:2} in Definition~\ref{defErrorState}): The shared output $AB!m$ departing from $v_0$ in $\mathcal{I}_A$ has a corresponding input $AB?m$ in the sequence from $u'_0$ in $\mathcal{I'}_B$,
however, the first transition from $u'_0$ is $AB?n$ which is another shared interface between $\mathcal{I}_A$ and $\mathcal{I'}_B$.  In other words, $\mathcal{I}_A$ wants to send $m$ to $\mathcal{I'}_B$ at state $v_0$, but $\mathcal{I'}_B$ will never receive $m$ from the state $u'_0$ since $m$ is not the expected message received at state $u'_0$, and $m$ blocks the buffer away
from $n$. 
$(3)$ The state $v_1u_1$ of $\mathcal{I}_{\{A,B\}}\otimes \mathcal{I}_C$ is an error state (by \eqref{def:2} in Definition~\ref{defErrorState}). The shared output $BC!y$ in $v_1$ has the corresponding input $BC?y$ in $u_1$, and the first transition from $u_1$ is $AC?x$, which is another shared interface between $\mathcal{I}_{\{A,B\}}$ and $\mathcal{I}_C$. However, the corresponding shared output $AC!x$ is not in each sequence from state $v_2$ in $\mathcal{I}_{\{A,B\}}$. In other words, $\mathcal{I}_{\{A,B\}}$ sends $y$ to $\mathcal{I}_C$ in $v_1$, then sends $m$ to $\mathcal{I}_C$ in $v_2$, but $y$ is not consumed by $\mathcal{I}_C$ since $\mathcal{I}_C$ wants to first receive $x$ from $\mathcal{I}_{\{A,B\}}$ in $u_1$, however, $x$ is not sent in $v_2$ in $\mathcal{I}_{\{A,B\}}$.


\begin{theorem}
Given two composable GIA $\mathcal{I}$ and $\mathcal{I'}$, an error state $(v,v')$ exists in $\mathcal{I}\otimes \mathcal{I'}$ if and only if at least a send action in the shared output departing from $v$ or $v'$ is never consumed.
\end{theorem}


\section{Group Interface Automata Based Verification}\label{sectGIAbV}

We advocate GIA for the verification of well-formedness of g-choreographies. According to \cite{gt16}, well-formedness of g-choreographies implies deadlock freedom. Now, we show that the well-formedness of g-choreographies is equivalent to the non-existence of error states and parallel, branching error states in the corresponding GIA. Therefore, non-existence of (parallel, branching) error states implies deadlock freedom. The following diagram describes our approach.\newline

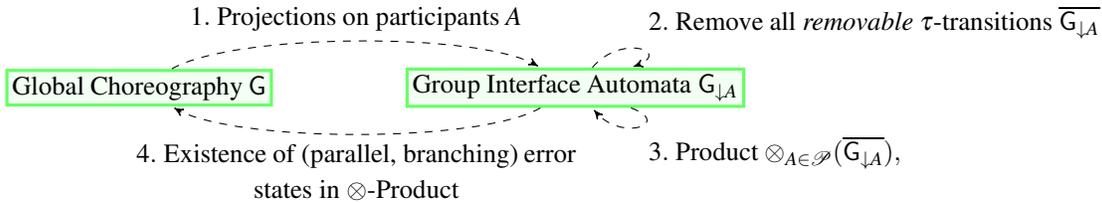
\begin{tikzpicture}[shorten >=1pt,node distance=5.8cm,>=stealth',auto,box/.style={rectangle, draw=green!60, fill=green!5, very thick, minimum size=7mm}]
\node[box,inner sep=2pt,minimum size=0pt]        (q_0)                           {\small \text{Global Choreography $\textsf{G}$}};  
\node[box,inner sep=2pt,minimum size=0pt]  (q_1)  [right of=q_0]     {\small \text{Group Interface Automata $\textsf{G}_{\downarrow A}$}}; 
\path[->]       
             (q_0)     edge  [dashed, out=30,in=150,looseness=0.5]  node{\small $1.$ Projections on participants $A$}     (q_1)
             (q_1)     edge  [dashed, out=210,in=330,looseness=0.5]  node[align=center]{\small $4.$ Existence of (parallel, branching) error \\  \small states  in $\otimes$-Product}     (q_0)
             (q_1)     edge  [dashed, out=50,in=20,looseness=3]  node{\small $2.$ Remove all \emph{removable} $\tau$-transitions $\overline{{\textsf{G}_{\downarrow A}}}$}     (q_1)
              (q_1)     edge  [dashed,  out=340,in=310,looseness=3]  node{\small $3.$ Product $\otimes_{A \in \mathcal{P}}(\overline{{\textsf{G}_{\downarrow A}}}),$}     (q_1);
\end{tikzpicture}

\medskip\noindent Roughly speaking, $(1)$ projection 
yields a set of GIA corresponding to each local participant involved
in the g-choreography; $(2)$  projection introduces
$\tau$-transitions, that is internal transitions that do not represent
communications; some of those $\tau$-transitions may lead to spurious
error states in the $\otimes$-product between GIA; 
the elimination of spurious $\tau$-transitions
reveals actual error states; 
$(3)$ we take the $\otimes$-product of all the GIA without removable $\tau$-transitions;
$(4)$ we analyse the GIA yielded by last step
to detect the
existence of error states, parallel and branching error states.



\subsection{Projection}\label{secProject}
We define some auxiliary notions before introducing the projection operation.
\begin{enumerate}
\item 
Let $\mathcal{I'}=(V', v_0', \mathcal{G'},  \mathcal{A'},\mathcal{T'})$ and $\mathcal{I''}=(V'', v_0'', \mathcal{G''}, \mathcal{A''},\mathcal{T''})$ be two $\mathcal G$-interface automata. We define $\mathcal{I'} \times \mathcal{I''}= (V' \times V'', (v_0', v_0''), \mathcal{G'}\cup \mathcal{G''}, \mathcal{A'}\cup \mathcal{A''}, \mathcal{T})$, where $((v',v''), \alpha, (u',u''))\in \mathcal{T}$ iff
$$((v', \alpha, u')\in \mathcal{T'} \ \text{and} \ v''=u''\in V'')\quad  \text{or} \quad ((v'', \alpha, u'')\in \mathcal{T''} \ \text{and} \ v'=u'\in V').$$

\item 
Let $\{u /v\} \mathcal{I}$ be the automaton obtained by substituting the state $v$ with the state $u$.

\item 
Let $\mathcal{I} \odot \textbf{n}$ be the automaton $(V \times \textbf{n}, (v_0, n), \mathcal{G}, \mathcal{A}, \mathcal{T} \odot \textbf{n})$, where $\mathcal{T} \odot \textbf{n}=\{((v,n), \alpha,(u,n))\mid (v,\alpha,u)\in \mathcal{T}\}$. 

\item 
Let $\mathcal{I'} \circ \mathcal{I''}$ be the automata $\mathcal{I}=(V' \cup V'', v_0', \mathcal{G'}\cup \mathcal{G''}, \mathcal{A'}\cup \mathcal{A''},\mathcal{T'} \cup \mathcal{T''})$, This operation is intended to help with connecting two automata by sequential or by branching.
\end{enumerate}

Then, we define a projection function by induction on the syntax of \textsf{G} returning a triple $(\mathcal{I}_A, v_0, v_e )$, where $\mathcal{I}_A$ is a GIA, $v_0$ is its initial state, and $v_e$ is the special state of $\mathcal{I}$ used to connect it to the other  GIA. Generally, the connecting state is the final state which does not have any leaving transitions. Here, we use $(\mathcal{I}_A, v_0, v_e) \odot \textbf{n}$ to represent $(\mathcal{I}_A \odot \textbf{n}, (v_0,n), (v_e,n))$.

  
\begin{definition}[Projection from g-choreographies to GIA]\label{defIAFromGG}
\textit{Let} $\textsf{G}$ \textit{be a g-choreography. The projections} $\textsf{G}_{\downarrow A}$ \textit{are defined as follows.}


\begin{eqnarray}
\textsf{G}_{{\downarrow}_A} =
\begin{cases}

\begin{tikzpicture}[shorten >=1pt,node distance=1.3cm,>=stealth',auto,
                                    every state/.style={thin,fill=blue!10}]
\draw (-0.8,-0.4) rectangle (2.2,0.4);
\node[state,initial,inner sep=2pt,minimum size=0pt]        (q_0)                           {\tiny $v_0$};  
\node[state,inner sep=2pt,minimum size=0pt]  (q_1)  [right of=q_0]     {\tiny $v_e$}; 
\path[->]       
             (q_0)     edge    node{\tiny $\tau$}     (q_1);
\end{tikzpicture}   & \parbox{10pc}{$\text{if} \  \textsf{G}={0}$\vspace{2em}}\\

\begin{tikzpicture}[shorten >=1pt,node distance=1.3cm,>=stealth',auto,
                                    every state/.style={thin,fill=blue!10}]
\draw (-0.8,-0.4) rectangle (2.2,0.4);
\node[state,initial,inner sep=2pt,minimum size=0pt]        (q_0)                           {\tiny $v_0$};  
\node[state,inner sep=2pt,minimum size=0pt]  (q_1)  [right of=q_0]     {\tiny $v_e$}; 
\path[->]       
             (q_0)     edge    node{\tiny $\tau$}     (q_1);
\end{tikzpicture}    & \parbox{10pc}{$\text{if} \  \textsf{G}=B\xrightarrow{m} C $\vspace{2em}}\\

\begin{tikzpicture}[shorten >=1pt,node distance=1.5cm,>=stealth',auto,
                                    every state/.style={thin,fill=blue!10}]
\draw (-0.8,-0.4) rectangle (2.3,0.4);
\node[state,initial,inner sep=2pt,minimum size=0pt]        (q_0)                           {\tiny $v_0$};  
\node[state,inner sep=2pt,minimum size=0pt]  (q_1)  [right of=q_0]     {\tiny $v_e$}; 
\draw  [->] (2.3,0.0)--(2.6,0.0);
\node at (2.75,-0.0) {\tiny $ABm$};
\path[->]       
             (q_0)     edge    node{\tiny $AB!m$}     (q_1);
\end{tikzpicture}  & \parbox{10pc}{$\text{if} \  \textsf{G}=A\xrightarrow{m} B \ \text{and} \ v_0\neq v_e$\vspace{2em}}\\

\begin{tikzpicture}[shorten >=1pt,node distance=1.5cm,>=stealth',auto,
                                    every state/.style={thin,fill=blue!10}]
\draw (-0.8,-0.4) rectangle (2.3,0.4);
\node[state,initial,inner sep=2pt,minimum size=0pt]        (q_0)                           {\tiny $v_0$};  
\node[state,inner sep=2pt,minimum size=0pt]  (q_1)  [right of=q_0]     {\tiny $v_e$}; 
\draw  [<-] (2.3,0.0)--(2.6,0.0);
\node at (2.75,-0.0) {\tiny $BAm$};
\path[->]       
             (q_0)     edge    node{\tiny $BA?m$}     (q_1);
\end{tikzpicture}  & \parbox{10pc}{$\text{if} \  \textsf{G}=B\xrightarrow{m} A \ \text{and} \ v_0\neq v_e$\vspace{2em}}\\

(\mathcal{I'}_{A} \circ \{v_e/u_0\}\mathcal{I''}_{A}, v_0,u_e)     & \text{if} \  \textsf{G}=\textsf{G'};\textsf{G''} \\
& \qquad \text{and} \ (\mathcal{I'}_{A}, v_0, v_e)=\textsf{G'}_{\downarrow A} \odot \textbf{1}\\
& \qquad \text{and} \ (\mathcal{I''}_{A}, u_0, u_e)=\textsf{G''}_{\downarrow A} \odot \textbf{2}\\

(\{u_e/v_e\}\mathcal{I'}_{A} \circ \{v_0/u_0\}\mathcal{I''}_{A}, v_0,u_e)      & \text{if} \  \textsf{G}=\textsf{G'}+\textsf{G''} \\
& \qquad \text{and} \ (\mathcal{I'}_{A}, v_0, v_e)=\textsf{G'}_{\downarrow A} \odot \textbf{1}\\
& \qquad \text{and} \ (\mathcal{I''}_{A}, u_0, u_e)=\textsf{G''}_{\downarrow A} \odot \textbf{2}\\

(\mathcal{I'}_{A} \times \mathcal{I''}_{A}, (v_0, u_0), (v_e, u_e))      & \text{if} \  \textsf{G}=\textsf{G'} \mid \textsf{G''} \\
& \qquad \text{and} \ (\mathcal{I'}_{A}, v_0, v_e,)=\overline{\textsf{G'}_{\downarrow A}} \odot \textbf{1}\\
& \qquad \text{and} \ (\mathcal{I''}_{A}, u_0, u_e)=\overline{\textsf{G''}_{\downarrow A}} \odot \textbf{2}\\
\end{cases}
\end{eqnarray}
\end{definition}

We inductively define the projections and preserve all the behaviours of each participant. Roughly speaking, ($1$) an empty g-choreography ${0}$ induces a GIA with two states connected by a $\tau$ transition, which are the initial state and the connecting state.  
($2$) An interaction $A\xrightarrow{m}B$ yields three different projections on the sender, receiver and other participants: Each projection has an initial state, a connecting state and a transition, where the transition action is an output (resp. input, $\tau$) if the projection is on the sender (resp.\ receiver, other participants). 
($3$) The projection ${(\textsf{G}'; \textsf{G}'')}_{\downarrow A}$ is obtained by merging the connecting state of ${\textsf{G}'}_{\downarrow A}$ with the initial state of ${\textsf{G}''}_{\downarrow A}$. 
($4$) The projection ${(\textsf{G}' + \textsf{G}'')}_{\downarrow A}$ is constructed by merging the initial states and the connecting states of ${\textsf{G}'}_{\downarrow A}$ and ${\textsf{G}''}_{\downarrow A}$. 
($5$) The projection ${(\textsf{G}' \mid \textsf{G}'')}_{\downarrow A}$ is generated by interleaving the transitions of $\overline{{\textsf{G}'}_{\downarrow A}}$ and $\overline{{\textsf{G}''}_{\downarrow A}}$, where $\overline{{\textsf{G}'}_{\downarrow A}}$ and $\overline{{\textsf{G}''}_{\downarrow A}}$ are the GIA without removable $\tau$-transitions refined from ${\textsf{G}'}_{\downarrow A}$ and ${\textsf{G}''}_{\downarrow A}$ respectively. Removable $\tau$-transitions will be introduced in next subsection.

\subsection{Removability of $\tau$-transitions}\label{sectRemTau}
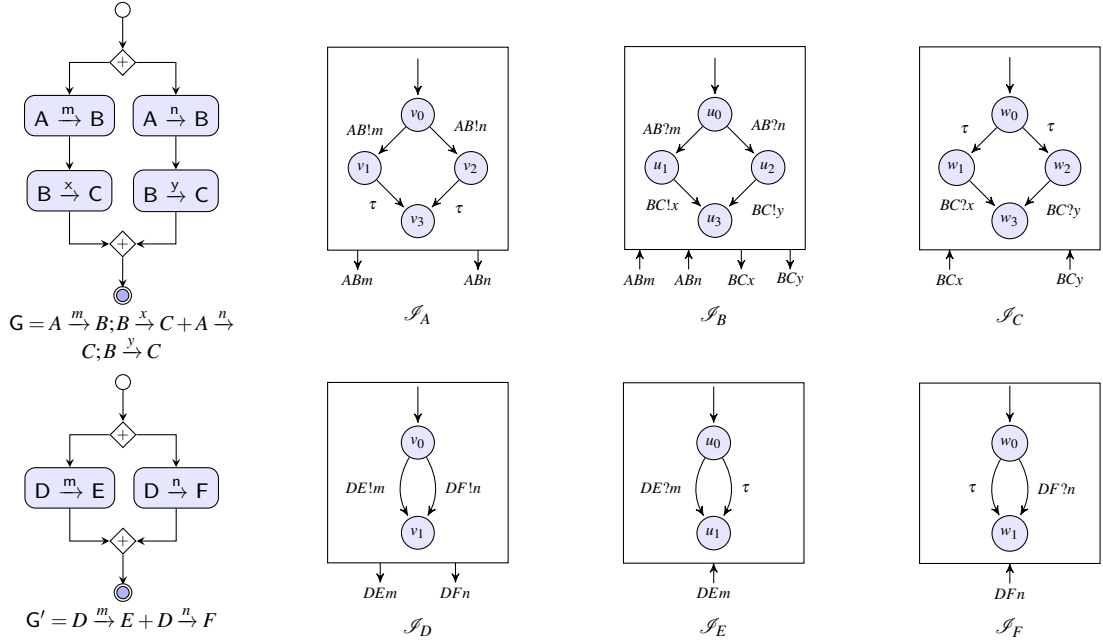
\begin{figure}[thbp]
\begin{minipage}{0.24\textwidth}
\centering
\begin{tikzpicture}[node distance=1cm,
    every node/.style={fill=white, font=\sffamily}, align=center]
  \node (0)             [start]              {};
  \node at (0, -0.7) (1)     [branch]          {\tiny $+$};
  \node (2)     [basic, below left of=1]          {\scriptsize A $\xrightarrow{\text{m}}$ B};
  \node (3)     [basic, below of=2]          {\scriptsize B $\xrightarrow{\text{x}}$ C};
  \node (4)      [basic, below right of=1]   {\scriptsize A $\xrightarrow{\text{n}}$ B};
  \node (5)     [basic, below of=4]          {\scriptsize B $\xrightarrow{\text{y}}$ C};
  \node (6)    [branch, below right of=3]              {\tiny $+$};
  \node at (0, -3.8) (7)    [end]              {};
  \draw[-{stealth}]      (0) -- (1);
  \draw[-{stealth}]      (1) -| (2);
  \draw[-{stealth}]      (2) -- (3);
  \draw[-{stealth}]      (1) -| (4);
  \draw[-{stealth}]      (4) -- (5);
  \draw[-{stealth}]      (3) |- (6);
  \draw[-{stealth}]      (5) |- (6);
  \draw[-{stealth}]      (6) -- (7);
\end{tikzpicture}
\centering
\\  \scriptsize $\textsf{G}=A\xrightarrow{m}B;B\xrightarrow{x}C+A\xrightarrow{n}C;B\xrightarrow{y}C$
\end{minipage}
\begin{minipage}{0.24\textwidth}
\begin{tikzpicture}[shorten >=1pt,node distance=1.0cm,>=stealth',auto,
                                    every state/.style={thin,fill=blue!10}]
\draw (-1.2,-1.8) rectangle (1.2,0.9);
\node at (-0.7,-0.2) {\tiny $AB!m$};
\node at (-0.6,-1.2) {\tiny $\tau$};
\draw  [->] (-0.8,-1.8)--(-0.8,-2.1);
\draw  [->] (0.8,-1.8)--(0.8,-2.1);
\node at (-0.8,-2.2) {\tiny $ABm$};
\node at (0.8,-2.2) {\tiny $ABn$};
\node[state,initial above,inner sep=2pt,minimum size=0pt]        (u_0)                           {\tiny $v_0$};  
\node[state,inner sep=2pt,minimum size=0pt]  (u_1)  [below left of=u_0]     {\tiny $v_1$}; 
\node[state,inner sep=2pt,minimum size=0pt]  (u_2)  [below right of=u_0]     {\tiny $v_2$};
\node[state,inner sep=2pt,minimum size=0pt]  (u_3)  [below  right of=u_1]     {\tiny $v_3$};
\path[->]       
             (u_0)     edge    node{}     (u_1)
                       edge    node{\tiny $AB!n$}     (u_2) 
             (u_1)     edge    node{\tiny }     (u_3)    
             (u_2)     edge    node{\tiny $\tau$}     (u_3);
\end{tikzpicture}
\centering
\\ \scriptsize $\mathcal{I}_{A}$
\end{minipage}
\begin{minipage}{0.24\textwidth}
\begin{tikzpicture}[shorten >=1pt,node distance=1.0cm,>=stealth',auto,
                                    every state/.style={thin,fill=blue!10}]
\draw (-1.2,-1.8) rectangle (1.2,0.9);
\node at (-0.7,-0.2) {\tiny $AB?m$};
\node at (-0.7,-1.2) {\tiny $BC!x$};
\draw  [<-] (-1.0,-1.8)--(-1.0,-2.1);
\draw  [->] (1.0,-1.8)--(1.0,-2.1);
\draw  [<-] (-0.35,-1.8)--(-0.35,-2.1);
\draw  [->] (0.35,-1.8)--(0.35,-2.1);
\node at (-1.0,-2.2) {\tiny $ABm$};
\node at (-0.35,-2.2) {\tiny $ABn$};
\node at (0.35,-2.2) {\tiny $BCx$};
\node at (1.0,-2.2) {\tiny $BCy$};
\node[state,initial above,inner sep=2pt,minimum size=0pt]        (u_0)                           {\tiny $u_0$};  
\node[state,inner sep=2pt,minimum size=0pt]  (u_1)  [below left of=u_0]     {\tiny $u_1$}; 
\node[state,inner sep=2pt,minimum size=0pt]  (u_2)  [below right of=u_0]     {\tiny $u_2$};
\node[state,inner sep=2pt,minimum size=0pt]  (u_3)  [below  right of=u_1]     {\tiny $u_3$};
\path[->]       
             (u_0)     edge    node{}     (u_1)
                       edge    node{\tiny $AB?n$}     (u_2) 
             (u_1)     edge    node{\tiny }     (u_3)    
             (u_2)     edge    node{\tiny $BC!y$}     (u_3);
\end{tikzpicture}
\centering
\\ \scriptsize $\mathcal{I}_{B}$
\end{minipage}
\begin{minipage}{0.24\textwidth}
\begin{tikzpicture}[shorten >=1pt,node distance=1.0cm,>=stealth',auto,
                                    every state/.style={thin,fill=blue!10}]
\draw (-1.2,-1.8) rectangle (1.2,0.9);
\node at (-0.6,-0.2) {\tiny $\tau$};
\node at (-0.7,-1.2) {\tiny $BC?x$};
\draw  [<-] (-0.8,-1.8)--(-0.8,-2.1);
\draw  [<-] (0.8,-1.8)--(0.8,-2.1);
\node at (-0.8,-2.2) {\tiny $BCx$};
\node at (0.8,-2.2) {\tiny $BCy$};
\node[state,initial above,inner sep=2pt,minimum size=0pt]        (u_0)                           {\tiny $w_0$};  
\node[state,inner sep=2pt,minimum size=0pt]  (u_1)  [below left of=u_0]     {\tiny $w_1$}; 
\node[state,inner sep=2pt,minimum size=0pt]  (u_2)  [below right of=u_0]     {\tiny $w_2$};
\node[state,inner sep=2pt,minimum size=0pt]  (u_3)  [below  right of=u_1]     {\tiny $w_3$};
\path[->]       
             (u_0)     edge    node{}     (u_1)
                       edge    node{\tiny $\tau$}     (u_2) 
             (u_1)     edge    node{\tiny }     (u_3)    
             (u_2)     edge    node{\tiny $BC?y$}     (u_3);
\end{tikzpicture}
\centering
\\ \scriptsize $\mathcal{I}_{C}$
\end{minipage}\newline

\begin{minipage}{0.24\textwidth}
\centering
\begin{tikzpicture}[node distance=1cm,
    every node/.style={fill=white, font=\sffamily}, align=center]
  \node (0)             [start]              {};
  \node at (0, -0.7) (1)     [branch]          {\tiny $+$};
  \node (2)     [basic, below left of=1]          {\scriptsize D $\xrightarrow{\text{m}}$ E};
  \node (3)      [basic, below right of=1]   {\scriptsize D $\xrightarrow{\text{n}}$ F};
  \node (4)    [branch, below right of=2]              {\tiny $+$};
  \node at (0, -2.8) (5)    [end]              {};
  \draw[-{stealth}]      (0) -- (1);
  \draw[-{stealth}]     (1) -| (2);
  \draw[-{stealth}]      (1) -| (3);
  \draw[-{stealth}]      (2) |- (4);
  \draw[-{stealth}]      (3) |- (4);
  \draw[-{stealth}]      (4) -- (5);
\end{tikzpicture}
\centering
\\  \scriptsize $\textsf{G}'=D\xrightarrow{m}E + D\xrightarrow{n}F$
\end{minipage}
\begin{minipage}{0.24\textwidth}
\begin{tikzpicture}[shorten >=1pt,node distance=1.2cm,>=stealth',auto,
                                    every state/.style={thin,fill=blue!10}]      
\draw (-1.2,-1.6) rectangle (1.2,0.8);
\draw  [->] (-0.5,-1.6)--(-0.5,-1.9);
\draw  [->] (0.5,-1.6)--(0.5,-1.9);
\node at (-0.5,-2.0) {\tiny $DEm$};                                             \node at (0.5,-2.0) {\tiny $DFn$};                                                             
\node at (-0.7,-0.6) {\tiny $DE!m$};                                                             
\node[state,initial above,inner sep=2pt,minimum size=0pt]        (q_0)                           {\tiny $v_0$};  
\node[state,inner sep=2pt,minimum size=0pt]  (q_1)  [below of=q_0]     {\tiny $v_1$}; 
\path[->]       
             (q_0)   edge  [bend right]  node{}     (q_1)  
                     edge  [bend left]  node{\tiny $DF!n$}     (q_1);
\end{tikzpicture}
\centering
\\ \scriptsize $\mathcal{I}_{D}$
\end{minipage}
\begin{minipage}{0.24\textwidth}
\begin{tikzpicture}[shorten >=1pt,node distance=1.2cm,>=stealth',auto,
                                    every state/.style={thin,fill=blue!10}]      
\draw (-1.2,-1.6) rectangle (1.2,0.8);
\draw  [<-] (-0.0,-1.6)--(-0.0,-1.9);
\node at (-0.0,-2.0) {\tiny $DEm$};                                             
\node at (-0.7,-0.6) {\tiny $DE?m$};                                                             
\node[state,initial above,inner sep=2pt,minimum size=0pt]        (q_0)                           {\tiny $u_0$};  
\node[state,inner sep=2pt,minimum size=0pt]  (q_1)  [below of=q_0]     {\tiny $u_1$}; 
\path[->]       
             (q_0)  edge  [bend right]  node{}     (q_1)   
                    edge  [bend left]  node{\tiny $\tau$}     (q_1);
\end{tikzpicture}
\centering
\\ \scriptsize $\mathcal{I}_{E}$
\end{minipage}
\begin{minipage}{0.24\textwidth}
\begin{tikzpicture}[shorten >=1pt,node distance=1.2cm,>=stealth',auto,
                                    every state/.style={thin,fill=blue!10}]      
\draw (-1.2,-1.6) rectangle (1.2,0.8);
\draw  [<-] (-0.0,-1.6)--(-0.0,-1.9);
\node at (-0.0,-2.0) {\tiny $DFn$};                                             
\node at (-0.5,-0.6) {\tiny $\tau$};                                                             
\node[state,initial above,inner sep=2pt,minimum size=0pt]        (q_0)                           {\tiny $w_0$};  
\node[state,inner sep=2pt,minimum size=0pt]  (q_1)  [below of=q_0]     {\tiny $w_1$}; 
\path[->]       
             (q_0)  edge  [bend right]  node{}     (q_1)   
                    edge  [bend left]  node{\tiny $DF?n$}     (q_1);
\end{tikzpicture}
\centering
\\ \scriptsize $\mathcal{I}_{F}$
\end{minipage}
\caption{Spurious error states}\label{fig:fakeErrSt}
\end{figure}

Projection introduces
$\tau$-transitions which induce error states in the $\otimes$-product. However, some error states may be spurious, that is, they do not correspond to deadlocks. For instance, Figure~\ref{fig:fakeErrSt} shows a
well-formed g-choreography $\textsf{G}$, a not well-formed g-choreography $\textsf{G'}$ and their projections.

The states $v_3u_1w_2$ and $v_3u_2w_1$ in
$\mathcal{I}_A\otimes \mathcal{I}_B \otimes \mathcal{I}_C$ and $v_0u_1w_1$, $v_0u_0w_1$, $v_0u_1w_0$ in $\mathcal{I}_D\otimes \mathcal{I}_E\otimes \mathcal{I}_F$ are error
states according to Definition~\ref{defErrorState}. However, the error states in $\mathcal{I}_A\otimes \mathcal{I}_B \otimes \mathcal{I}_C$ are
spurious error states, whilst the error states in $\mathcal{I}_D\otimes \mathcal{I}_E\otimes \mathcal{I}_F$ do reflect deadlocks in $\textsf{G'}$. Therefore, in order to
disingiush spurious error states from error states, some $\tau$-transitions will be accounted for 
as removable. 
%
%
We first define the language between two states in GIA. We
use $\alpha, \beta$ to range over letters and $\omega$ to range
over words; as usual, the concatenation of words $\omega$ and
$\omega'$ is written as $\omega \omega'$ and $\tau$ is the
neutral element of word concatenation. 
Given a GIA
$\mathcal{I}=(V,v_0, \mathcal{G}, \mathcal{A}, \mathcal{T})$, the
language
between two states $v,v'\in V$ is
\[
\mathcal{L}(v,v')=
\begin{cases}
\{\alpha_0\cdots\alpha_n\mid v\xrightarrow{\alpha_0}v_1 \xrightarrow{\alpha_1}\cdots \xrightarrow{\alpha_{n}}v'\}  &\textit{ if $v\neq v'$}\\
\{\alpha_0\cdots\alpha_n\mid v\xrightarrow{\alpha_0}v_1 \xrightarrow{\alpha_1}\cdots \xrightarrow{\alpha_{n}}v'\}\cup\{\tau\}    &\textit{ if $v=v'$.}
\end{cases}
\]

Next, we give the definition of removability of $\tau$-transitions.

\begin{definition}[Removable $\tau$-transitions]\label{remtauIA}\it
  Given a GIA
  $\mathcal{I}=(V,v_0, \mathcal{G}, \mathcal{A},\mathcal{T})$, a
  $\tau$-transition $v\xrightarrow{\tau} v_{\tau}\in \mathcal{T}$ is
  removable if, for all $v'\in V$,
  $\mathcal{L}(v,v')\neq \emptyset$ and for all
  $\alpha \omega \in \mathcal{L}(v,v')$ we have
  \[
    \alpha\omega\neq \tau \implies \exists v_\tau'\in V: \mathcal{L}(v_{\tau}, v_\tau')=\{\tau\} \ \land\  v' \ \alpha\omega\text{-compatible to } v_\tau'
  \]
  where $v'$ is $\alpha\omega$-compatible to $v_\tau'$ if for all
  $\bar v \in V$ such that
  $\mathcal{L}(v_\tau', \bar v) \neq \{\tau\} \land \mathcal{L}(v_\tau', \bar v)
  \neq \emptyset$
  \begin{enumerate}
  \item\label{remCase2} if there is a sequence in
    $\mathcal{L}(v_\tau', \hat v)$ starting with a communication action
    different than $\alpha$ then for all
    $\beta\omega' \in \mathcal{L}(v_\tau', \hat v)$ such that
    $\beta \neq \alpha$ and either both $\beta$ and
    $\alpha$ are output actions or they are both input actions;
  \item\label{remCase1} otherwise, for all $\hat v \in V$ reachable
    from $\bar v$ with some communication transitions, then
    \[
      \mathcal{L}(v_\tau', \bar v) \mathcal{L}(\bar v, \hat v) =
      \{\alpha\omega\}
      \qquad\text{and}\qquad
      \mathcal{L}(\hat v, v') \in \{\emptyset, \{ \tau \}\}
      \ni
      \mathcal{L}(v', \hat v)
    \]
  \end{enumerate}
\end{definition}

Roughly speaking, a $\tau$-transition $v\xrightarrow{\tau}v_{\tau}$ is removable when either of the two following conditions are satisfied. $(1)$ The $\tau$-transition $v\xrightarrow{\tau}v_{\tau}$ 
does not affect the non-deterministic choices from the state $v$. $(2)$ The first non $\tau$-transition actions leaving from states $v$ and $v_{\tau}$ are all input interfaces or all output interfaces. 

\begin{figure}[htbp]
\begin{minipage}{0.33\textwidth}
\begin{tikzpicture}[shorten >=1pt,node distance=1.0cm,>=stealth',auto,
                                    every state/.style={thin,fill=blue!10}] 
\draw (-1.2,-1.8) rectangle (1.2,0.8);   
\draw  [->] (-0.0,-1.8)--(-0.0,-2.2);
\draw  [<-] (0.0,-1.8)--(0.0,-2.2);
\node at (0.0,-2.3) {\tiny $\alpha$};
\node at (-0.6,-0.2) {\color{red}\tiny $\tau$};     
\node at (-0.6,-1.2) {\tiny $\alpha$};                                                                                
\node[state,initial above,inner sep=2pt,minimum size=0pt]        (q_0)                           {\tiny $v_{}$};  
\node[state,inner sep=2pt,minimum size=0pt]  (q_1)  [ below left of=q_0]     {\tiny $v_{\tau}$}; 
\node[state,inner sep=2pt,minimum size=0pt]  (q_2)  [ below right of=q_0]     {\tiny $v'_{\tau}$}; 
\node[state,inner sep=2pt,minimum size=0pt]  (q_3)  [ below right of=q_1]     {\tiny $v_\tau'$}; 

\path[->]       
             (q_0)     edge [draw=red]  node{}     (q_1)           
             (q_0)     edge [draw=red]  node{\color{red}\tiny $\tau$}     (q_2)           
             (q_1)     edge   node{}     (q_3)           
             (q_2)     edge   node{\tiny $\alpha$}     (q_3)           
             ;
\end{tikzpicture}
\centering
\\ \scriptsize $\mathcal{I}_A$
\end{minipage}
\begin{minipage}{0.33\textwidth}
\begin{tikzpicture}[shorten >=1pt,node distance=1.0cm,>=stealth',auto,
                                    every state/.style={thin,fill=blue!10}] 
\draw (-1.2,-1.8) rectangle (1.2,0.8);   
\draw  [->] (-0.5,-1.8)--(-0.5,-2.2);
\draw  [<-] (-0.5,-1.8)--(-0.5,-2.2);
\draw  [->] (0.5,-1.8)--(0.5,-2.2);
\draw  [<-] (0.5,-1.8)--(0.5,-2.2);
\node at (-0.5,-2.3) {\tiny $\alpha$};
\node at (0.5,-2.3) {\tiny $\beta$};
\node at (-0.6,-0.2) {\color{red}\tiny $\tau$};     
\node at (-0.6,-1.2) {\tiny $\alpha$};                                                                                         
\node[state,initial above,inner sep=2pt,minimum size=0pt]        (q_0)                           {\tiny $v_{}$};  
\node[state,inner sep=2pt,minimum size=0pt]  (q_1)  [ below left of=q_0]     {\tiny $v_{\tau}$}; 
\node[state,inner sep=2pt,minimum size=0pt]  (q_2)  [ below right of=q_0]     {\tiny $v_{\beta}$}; 
\node[state,inner sep=2pt,minimum size=0pt]  (q_3)  [ below right of=q_1]     {\tiny $v_\tau'$}; 

\path[->]       
             (q_0)     edge [draw=red]  node{}     (q_1)           
             (q_0)     edge   node{\tiny $\beta$}     (q_2)           
             (q_1)     edge   node{}     (q_3)           
             (q_2)     edge [draw=red]  node{\color{red}\tiny $\tau$}     (q_3)           
             ;
\end{tikzpicture}
\centering
\\ \scriptsize $\mathcal{I}_B$
\end{minipage}
\begin{minipage}{0.33\textwidth}
\begin{tikzpicture}[shorten >=1pt,node distance=1.1cm,>=stealth',auto,
                                    every state/.style={thin,fill=blue!10}]
\draw (-1.0,-1.5) rectangle (1.0,0.8);
\draw  [<-] (0.0,-1.5)--(0.0,-1.9);
\draw  [->] (0.0,-1.5)--(0.0,-1.9);
\node at (0.0,-2.0) {\tiny $\alpha$};                                                             
\node at (-0.6,-0.6) {\tiny $\tau$};                                                                  
\node[state,initial above,inner sep=2pt,minimum size=0pt]        (q_0)                           {\tiny $v$};  
\node[state,inner sep=2pt,minimum size=0pt]  (q_1)  [below of=q_0]     {\tiny $v_\tau$}; 

\path[->]       
             (q_0)     edge  [bend right]  node{}     (q_1)
                       edge  [bend left]  node{\tiny $\alpha$}     (q_1);
\end{tikzpicture}
\centering
\\ \tiny $\mathcal{I}_{C}$
\end{minipage}
\caption{Removability of $\tau$-transitions } \label{fig:removeable0}
\end{figure}
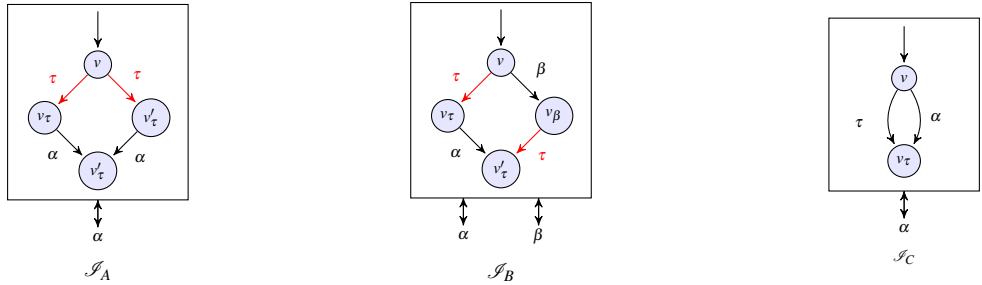

For instance, Figure~\ref{fig:removeable0} shows three simple instances of the removability of $\tau$-transitions, the first two machines reflect case~\eqref{remCase1} and case~\eqref{remCase2} respectively. 
In the GIA $\mathcal{I}_A$, the $\tau$-transitions $v_{}\xrightarrow{\tau}v_{\tau}$ and $v_{}\xrightarrow{\tau}v'_{\tau}$ denoted by red colour are removable. In the GIA $\mathcal{I}_B$, $v_{\beta}\xrightarrow{\tau}v_\tau'$ 
is removable, and if $\alpha, \beta$ are both inputs or both outputs, then the $\tau$-transition $v\xrightarrow{\tau}v_{\tau}$ 
is removable (and otherwise it is not). 
In the GIA $\mathcal{I}_C$, 
$v\xrightarrow{\tau}v_{\tau}$ is not removable since 
it affects the choices from $v$.

After finding all removable $\tau$-transitions in projections, we construct the GIA without removable $\tau$-transitions from $\tau$-equivalence classes.


\begin{definition}[$\tau$-Equivalence class of states]
  \textit{Let $\simeq_\tau$ be the smallest equivalence relation containing all $v\simeq_\tau v'$, where $v\xrightarrow{\tau}\cdots \xrightarrow{\tau}v'$ or $v’\xrightarrow{\tau}\cdots \xrightarrow{\tau}v$, and each $\tau$-transition is removable. We write $[v]_{\simeq_\tau}$ as the equivalence class of the state $v$ w.r.t. $\simeq_\tau$.}
\end{definition}

We define  $[V]_{\simeq_\tau}=\bigcup_{v\in V}[v]_{\simeq_\tau}$, where $V$ is a set of states. 

\begin{definition}\label{IAwithoutTau}
\textit{Let $\mathcal{I}=(V, v_0, \mathcal{G}, \mathcal{A}, \mathcal{T})$ be a GIA. Then $\overline{\mathcal{I}}=(\overline{V}, \overline{v_0}, \overline{\mathcal{G}}, \overline{\mathcal{A}},\overline{\mathcal{T}})$ is the GIA without removable $\tau$-transitions constructed from $\mathcal{I}$, where}
$$\overline{V} = \{ [v]_{\simeq_\tau} \mid v\in V\}\cup \{\emptyset\}, \quad   \overline{v_0}= [v_0]_{\simeq_\tau},\quad \overline{\mathcal{G}}=\mathcal{G}, \quad \overline{\mathcal{A}}=\mathcal{A}, \quad \overline{\mathcal{T}}=\{([v]_{\simeq_\tau}, \alpha, [v']_{\simeq_\tau})\mid (v,\alpha,v')\in \mathcal{T}\}.$$
\end{definition}

\begin{theorem}\label{langEqOfIA}
Let $\textsf{G}$ be a g-choreography, $
\textsf{G}_{\downarrow A}=\mathcal{I}_A=(V, v_0,\{A\},\mathcal{A}, \mathcal{T})$ be a GIA projected from participant $A$ where $A\in \mathcal{P}$. Let $\overline{\textsf{G}_{\downarrow A}}=\overline{\mathcal{I}_A}=(\overline{V}, \overline{v_0}, \{A\},\overline{\mathcal{A}},\overline{\mathcal{T}})$ be the GIA without removable $\tau$-transitions constructed from $\textsf{G}_{\downarrow A}$. Then $\textsf{G}_{\downarrow A}$ and $\overline{\textsf{G}_{\downarrow A}}$ are language equivalent. 
\end{theorem}

\begin{figure}[htbp]
\begin{minipage}{0.22\textwidth}
\begin{tikzpicture}[shorten >=1pt,node distance=1.0cm,>=stealth',auto,
                                    every state/.style={thin,fill=blue!10}]
\draw (-1.1,-1.5) rectangle (1.1,0.9);
\node at (-0.7,-0.5) {\tiny $AB!m$};
\draw  [->] (-0.8,-1.5)--(-0.8,-1.8);
\draw  [->] (0.8,-1.5)--(0.8,-1.8);
\node at (-0.8,-1.9) {\tiny $ABm$};
\node at (0.8,-1.9) {\tiny $ABn$};
\node[state,initial above,inner sep=2pt,minimum size=0pt]        (u_0)                           {\tiny $v_0$};  
\node[state,inner sep=2pt,minimum size=0pt]  (u_1)  [below of=u_0]     {\tiny $v_1$}; 
\path[->]       
             (u_0)     edge [bend right]    node{}     (u_1)
                       edge [bend left]    node{\tiny $AB!n$}     (u_1);
\end{tikzpicture}
\centering
\\ \scriptsize $\overline{\mathcal{I}_{A}}$
\end{minipage}
\begin{minipage}{0.22\textwidth}
\begin{tikzpicture}[shorten >=1pt,node distance=1.0cm,>=stealth',auto,
                                    every state/.style={thin,fill=blue!10}]
\draw (-1.2,-1.8) rectangle (1.2,0.9);
\node at (-0.7,-0.2) {\tiny $AB?m$};
\node at (-0.7,-1.2) {\tiny $BC!x$};
\draw  [<-] (-1.0,-1.8)--(-1.0,-2.1);
\draw  [->] (1.0,-1.8)--(1.0,-2.1);
\draw  [<-] (-0.35,-1.8)--(-0.35,-2.1);
\draw  [->] (0.35,-1.8)--(0.35,-2.1);
\node at (-1.0,-2.2) {\tiny $ABm$};
\node at (-0.35,-2.2) {\tiny $ABn$};
\node at (0.35,-2.2) {\tiny $BCx$};
\node at (1.0,-2.2) {\tiny $BCy$};
\node[state,initial above,inner sep=2pt,minimum size=0pt]        (u_0)                           {\tiny $u_0$};  
\node[state,inner sep=2pt,minimum size=0pt]  (u_1)  [below left of=u_0]     {\tiny $u_1$}; 
\node[state,inner sep=2pt,minimum size=0pt]  (u_2)  [below right of=u_0]     {\tiny $u_2$};
\node[state,inner sep=2pt,minimum size=0pt]  (u_3)  [below  right of=u_1]     {\tiny $u_3$};
\path[->]       
             (u_0)     edge    node{}     (u_1)
                       edge    node{\tiny $AB?n$}     (u_2) 
             (u_1)     edge    node{\tiny }     (u_3)    
             (u_2)     edge    node{\tiny $BC!y$}     (u_3);
\end{tikzpicture}
\centering
\\ \scriptsize $\overline{\mathcal{I}_{B}}$
\end{minipage}
\begin{minipage}{0.22\textwidth}
\begin{tikzpicture}[shorten >=1pt,node distance=1.0cm,>=stealth',auto,
                                    every state/.style={thin,fill=blue!10}]
\draw (-1.1,-1.5) rectangle (1.1,0.9);
\node at (-0.7,-0.5) {\tiny $BC?x$};
\draw  [<-] (-0.8,-1.5)--(-0.8,-1.8);
\draw  [<-] (0.8,-1.5)--(0.8,-1.8);
\node at (-0.8,-1.9) {\tiny $BCx$};
\node at (0.8,-1.9) {\tiny $BCy$};
\node[state,initial above,inner sep=2pt,minimum size=0pt]        (u_0)                           {\tiny $w_0$};  
\node[state,inner sep=2pt,minimum size=0pt]  (u_1)  [below of=u_0]     {\tiny $w_1$}; 
\path[->]       
             (u_0)     edge [bend right]   node{}     (u_1)
                       edge [bend left]   node{\tiny $BC?y$}     (u_1);
\end{tikzpicture}
\centering
\\ \scriptsize $\overline{\mathcal{I}_{C}}$
\end{minipage}
\begin{minipage}{0.30\textwidth}
\begin{tikzpicture}[shorten >=1pt,node distance=1.3cm,>=stealth',auto,
                                    every state/.style={thin,fill=blue!10}, dia/.style={diamond, draw=green!60, very thick, fill=green!10}]      
\draw (-1.7,-2.5) rectangle (1.7,1.1);
\node at (-1.0,-0.3) {\tiny $AB!?m$};                                          
\node at (-1.0,-1.8) {\tiny $BC!?x$};                                    
\node[state,initial above,inner sep=2pt,minimum size=0pt]        (q_0)                           {\tiny $v_0u_0w_0$};  
\node[state,inner sep=2pt,minimum size=0pt]  (q_1)  [below left of=q_0]     {\tiny $v_1u_1w_0$}; 
\node[state,inner sep=2pt,minimum size=0pt]  (q_2)  [below right of=q_0]     {\tiny $v_1u_2w_0$}; 
\node[state,inner sep=2pt,minimum size=0pt]  (q_3)  [below right of=q_1]     {\tiny $v_1u_3w_1$}; 
\path[->]       
             (q_0)    edge    node{}     (q_1)            		 
             (q_0)    edge    node{\tiny $AB!?n$}     (q_2)
             (q_1)    edge []   node{}     (q_3) 
             (q_2)    edge []   node{\tiny $BC!?y$}     (q_3);
\end{tikzpicture}
\centering 
\\ \scriptsize $\overline{\mathcal{I}_A} \otimes \overline{\mathcal{I}_B} \otimes \overline{\mathcal{I}_C}$
\end{minipage}
\caption{Refined GIA and their $\otimes$-product}\label{fig:verfWf}
\end{figure}
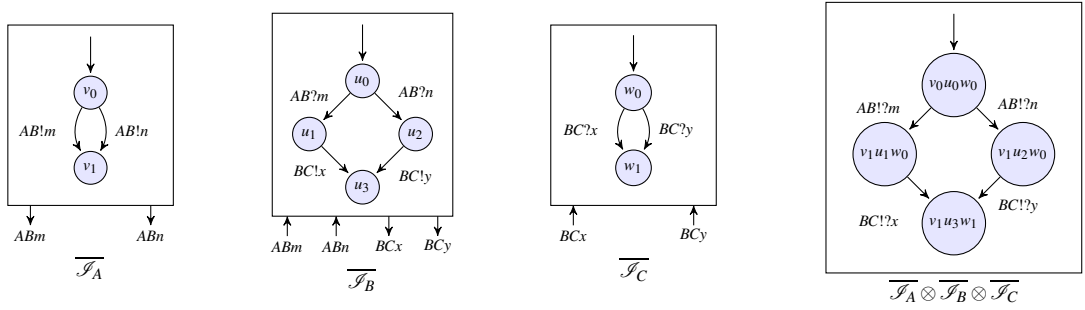

For instance, Figure~\ref{fig:verfWf} shows the GIA with without removable $\tau$-transitions $\overline{\mathcal{I}_A}$, $\overline{\mathcal{I}_B}$ and $\overline{\mathcal{I}_C}$ refined from $\textsf{G}$ in Figure~\ref{fig:fakeErrSt}, as well as their $\otimes$-product $\overline{\mathcal{I}_A}\otimes \overline{\mathcal{I}_B} \otimes \overline{\mathcal{I}_C}$ which does not have any error states.


\subsection{Parallel and branching error state}\label{secBranError}
In GIA based verification, we need to supplement error states with
new concepts called parallel error states and branching error states. As we have seen, error
states capture deadlocks. However, the non-existence of error states
is not sufficient to guarantee the well-formedness of parallel and branching
compositions. This leads to the definition of parallel and branching error
states. Again, we need some auxiliary notation first. We extend the
definition of $\textit{sbj}$ and $\textit{obj}$ from
Def.\ref{defLposet} to sequences in the obvious way
\[
  \textit{sbj}(\{\omega\mid \omega=\alpha_0\alpha_1\cdots\alpha_n\})=\bigcup_{0<i<n}\{\textit{sbj}(\alpha_i)\},
  \quad
  \textit{obj}(\{\omega\mid \omega=\alpha_0\alpha_1\cdots\alpha_n\})=\bigcup_{0<i<n}\{\textit{obj}(\alpha_i)\}
\]
also, we set
$\textit{sobj}(AB!?m)=\{A,B\}$ and define
$$\textit{sobj}(\{\alpha_0\alpha_1\cdots\alpha_n\})=\textit{sbj}(\{\alpha_0\alpha_1\cdots\alpha_n\})\cup \textit{obj}(\{\alpha_0\alpha_1\cdots\alpha_n\}).$$
$$\textit{sobj}(\{\omega\mid \omega=\alpha_0\alpha_1\cdots\alpha_n\})=\bigcup_{0<i<n}\textit{sobj}(\alpha_i)$$

\begin{definition}[Parallel error states]\label{parDefErrorState}
  \textit{Given a parallel composition}
  $\textsf{G}=\textsf{G'}\mid\textsf{G''}$, \textit{state $v$ is a
    parallel error state in}
  $\otimes_{A\in\mathcal{P}}\overline{{(\textsf{G'}\mid\textsf{G''})}_{\downarrow
      A}}=(V,v_0,\mathcal{G}, \mathcal{A}, \mathcal{T})$ \textit{if}
  $$\exists v \xrightarrow{\alpha} u\xrightarrow{\alpha}w \in
      \mathcal{T}, v\xrightarrow{\alpha}
      u'\xrightarrow{\alpha}w \in \mathcal{T} : u\neq u' \land
      \alpha \neq \tau.$$
  \textit{Let}
  $parError(\otimes_{A\in\mathcal{P}}\overline{{(\textsf{G'}\mid\textsf{G''})}_{\downarrow
      A}})$ \textit{be the set of parallel error states of}
  $\otimes_{A\in\mathcal{P}}\overline{{(\textsf{G'}\mid\textsf{G''})}_{\downarrow
      A}}$.
\end{definition}

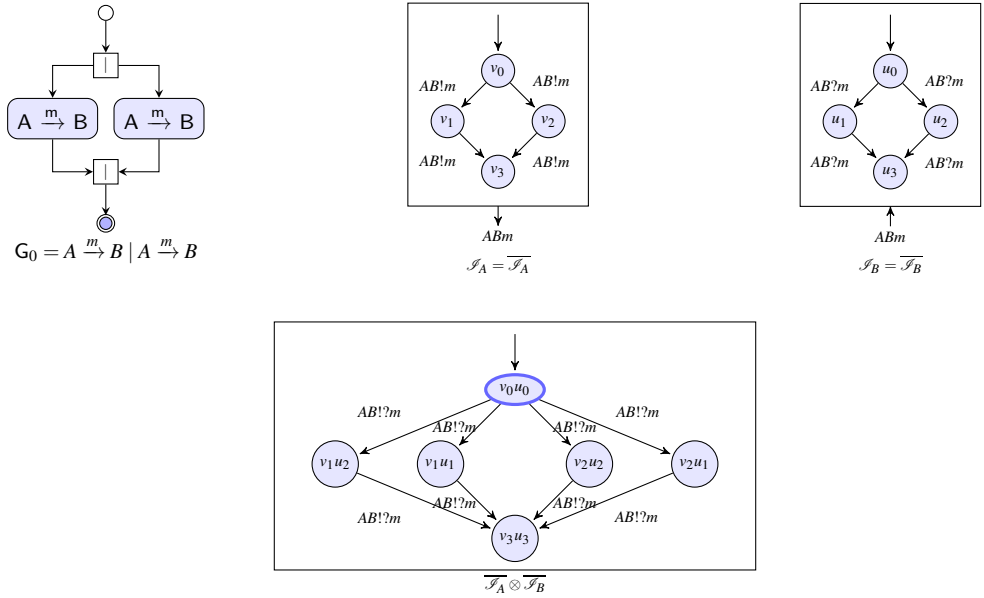
\begin{figure}[h]
  \begin{minipage}{0.32\textwidth}
    \centering
    \begin{tikzpicture}[node distance=1cm,
    every node/.style={fill=white, font=\sffamily}, align=center]
  \node (0)             [start]              {};
  \node at (0, -0.7) (1)     [folk]          {\tiny $|$};
  \node (2)     [basic, below left of=1]          {\scriptsize A $\xrightarrow{\text{m}}$ B};
  \node (3)      [basic, below right of=1]   {\scriptsize A $\xrightarrow{\text{m}}$ B};
  \node (4)    [folk, below right of=2]              {\tiny $|$};
  \node at (0, -2.8) (5)    [end]              {};
  \draw[-{stealth}]      (0) -- (1);
  \draw[-{stealth}]     (1) -| (2);
  \draw[-{stealth}]      (1) -| (3);
  \draw[-{stealth}]      (2) |- (4);
  \draw[-{stealth}]      (3) |- (4);
  \draw[-{stealth}]      (4) -- (5);
\end{tikzpicture}
    \centering
    \\  \scriptsize $\textsf{G}_0=A\xrightarrow{m}B \mid A\xrightarrow{m}B$
  \end{minipage}
  \begin{minipage}{0.32\textwidth}
    \begin{tikzpicture}[shorten >=1pt,node distance=0.95cm,>=stealth',auto,
      every state/.style={thin,fill=blue!10}]
      \draw (-1.2,-1.8) rectangle (1.2,0.9);
      \node at (-0.8,-0.2) {\tiny $AB!m$};
      \node at (-0.8,-1.2) {\tiny $AB!m$};
      \draw  [->] (-0.0,-1.8)--(-0.0,-2.1);
      \node at (-0.0,-2.2) {\tiny $ABm$};
      \node[state,initial above,inner sep=2pt,minimum size=0pt]        (u_0)                           {\tiny $v_0$};  
      \node[state,inner sep=2pt,minimum size=0pt]  (u_1)  [below left of=u_0]     {\tiny $v_1$}; 
      \node[state,inner sep=2pt,minimum size=0pt]  (u_2)  [below right of=u_0]     {\tiny $v_2$};
      \node[state,inner sep=2pt,minimum size=0pt]  (u_3)  [below right of=u_1]     {\tiny $v_3$};
      \path[->]       
      (u_0)     edge    node{}     (u_1)
      edge    node{\tiny $AB!m$}     (u_2) 
      (u_1)     edge    node{}     (u_3)
      (u_2)     edge    node{\tiny $AB!m$}     (u_3);
    \end{tikzpicture}
    \centering
    \\ \tiny $\mathcal{I}_{ A}=\overline{\mathcal{I}_{ A}}$
  \end{minipage}
  \begin{minipage}{0.32\textwidth}
    \begin{tikzpicture}[shorten >=1pt,node distance=0.95cm,>=stealth',auto,
      every state/.style={thin,fill=blue!10}]
      \draw (-1.2,-1.8) rectangle (1.2,0.9);
      \node at (-0.8,-0.2) {\tiny $AB?m$};
      \node at (-0.8,-1.2) {\tiny $AB?m$};
      \draw  [<-] (-0.0,-1.8)--(-0.0,-2.1);
      \node at (-0.0,-2.2) {\tiny $ABm$};
      \node[state,initial above,inner sep=2pt,minimum size=0pt]        (u_0)                           {\tiny $u_0$};  
      \node[state,inner sep=2pt,minimum size=0pt]  (u_1)  [below left of=u_0]     {\tiny $u_1$}; 
      \node[state,inner sep=2pt,minimum size=0pt]  (u_2)  [below right of=u_0]     {\tiny $u_2$};
      \node[state,inner sep=2pt,minimum size=0pt]  (u_3)  [below right of=u_1]     {\tiny $u_3$};
      \path[->]       
      (u_0)     edge    node{}     (u_1)
      edge    node{\tiny $AB?m$}     (u_2) 
      (u_1)     edge    node{}     (u_3)
      (u_2)     edge    node{\tiny $AB?m$}     (u_3);
    \end{tikzpicture}
    \centering
    \\ \tiny $\mathcal{I}_{ B}=\overline{\mathcal{I}_{ B}}$
  \end{minipage}\newline\newline
  
  \begin{minipage}{1.0\textwidth}
    \begin{tikzpicture}[shorten >=1pt,node distance=1.4cm,>=stealth',auto,
      every state/.style={thin,fill=blue!10}, ell/.style={ellipse, draw=blue!60, very thick, fill=blue!10}]
      \draw (-3.2,-2.4) rectangle (3.2,0.9);
      \node at (-1.8,-0.3) {\tiny $AB!?m$};
      \node at (-1.8,-1.7) {\tiny $AB!?m$};
      \node at (-0.8,-0.5) {\tiny $AB!?m$};
      \node at (0.8,-0.5) {\tiny $AB!?m$};
      \node at (-0.8,-1.5) {\tiny $AB!?m$};
      \node at (0.8,-1.5) {\tiny $AB!?m$};
      \node[ell,initial above,inner sep=2pt,minimum size=0pt]        (u_0)                           {\tiny $v_0u_0$};  
      \node[state,inner sep=2pt,minimum size=0pt]  (u_1)  [below left of=u_0]     {\tiny $v_1u_1$}; 
      \node[state,inner sep=2pt,minimum size=0pt]  (u_2)  [below right of=u_0]     {\tiny $v_2u_2$};
      \node[state,inner sep=2pt,minimum size=0pt]  (u_3)  [below right of=u_1]     {\tiny $v_3u_3$};
      \node[state,inner sep=2pt,minimum size=0pt]  (u_4)  [left of=u_1]     {\tiny $v_1u_2$};
      \node[state,inner sep=2pt,minimum size=0pt]  (u_5)  [right of=u_2]     {\tiny $v_2u_1$};
      \path[->]       
      (u_0)     edge    node{}     (u_1)
      edge    node{}     (u_2)
      edge    node{}     (u_4)
      edge    node{\tiny $AB!?m$}     (u_5) 
      (u_1)     edge    node{}     (u_3)
      (u_2)     edge    node{}     (u_3)
      (u_4)     edge    node{}     (u_3)
      (u_5)     edge    node{\tiny $AB!?m$}     (u_3);
    \end{tikzpicture}
    \centering
    \\ \tiny $\overline{\mathcal{I}_{ A}}\otimes \overline{\mathcal{I}_{ B}}$
  \end{minipage}
  
  \caption{Parallel error state}\label{fig:parnoten}
\end{figure}

Parallel error states capture the non well-formedness of parellel
g-choreographies that cannot be detected by error states. According to
Definition~\ref{genWf}, we know that well-forkedness needs there
is no common interactions existing in two threads. And according to
Definition~\ref{defIAFromGG}, we know that the projection of parallel
compositions is constructed by interleaving the transitions from two
threads. Roughly, if the product generates a state $v$ with 
transitions with the same label $\alpha$ that form a diamond, then $v$ is a
parallel error state. For instance, in Figure~\ref{fig:parnoten}, the
parallel composition $\textsf{G}_0$ is not well-forked. There are no error states in
$\overline{\mathcal{I}_{ A}}\otimes \overline{\mathcal{I}_{ B}}$ (cf. Figure~\ref{fig:parnoten}), but
the initial state $v_0u_0$ of
$\overline{\mathcal{I}_{ A}}\otimes \overline{\mathcal{I}_{ B}}$ is a
parallel error states.

\begin{definition}[Branching error states]\label{branchingDefErrorState}
  \textit{Given a branching composition}
  $\textsf{G}=\textsf{G'}+\textsf{G''}$, \textit{we say states $v$ and $v'$ ($v\neq v'$)  
    are branching error states in}
  $\otimes_{A\in\mathcal{P}}\overline{{(\textsf{G'}+\textsf{G''})}_{\downarrow
      A}}=(V,v_0,\mathcal{G}, \mathcal{A}, \mathcal{T}),$ \textit{if
    either of the following conditions holds}
  \begin{enumerate}
  \item\label{brdef:1}
    $
    v=v_0 \land |\mathcal{T}|=1$,
    
  \item\label{brdef:2}
    $
    \mathcal{L}(v_0,v)=\{\tau\}\land (\exists (v\xrightarrow{AB!?m}u\land v\xrightarrow{CD!?n}w): u\neq w \land (A\neq C \lor AB!?m=CD!?n)$), 

  \item\label{brdef:3}  
    $\displaystyle\begin{aligned}[t]
      (\mathcal{L}(v_0,v)\neq \{\tau\}\lor \mathcal{L}(v_0,v')\neq \{\tau\})&\land (\exists (v\xrightarrow{AB!?m}u \land v'\xrightarrow{AB!?m}u') : \\ 
      &(A\notin \textit{sobj}(\mathcal{L}(v_0,v))
      \land A\notin \textit{sobj}(\mathcal{L}(v_0,v')))\lor\\
      &(B\notin \textit{sobj}(\mathcal{L}(v_0,v))
      \land B\notin\textit{sobj}(\mathcal{L}(v_0,v')))).   
    \end{aligned}$
  \end{enumerate}
  \textit{Let} $brcError(\otimes_{A\in\mathcal{P}}\overline{{(\textsf{G'}+\textsf{G''})}_{\downarrow A}})$ \textit{be the set of branching error states of} $\otimes_{A\in\mathcal{P}}\overline{{(\textsf{G'}+\textsf{G''})}_{\downarrow A}}$.
  
\end{definition}

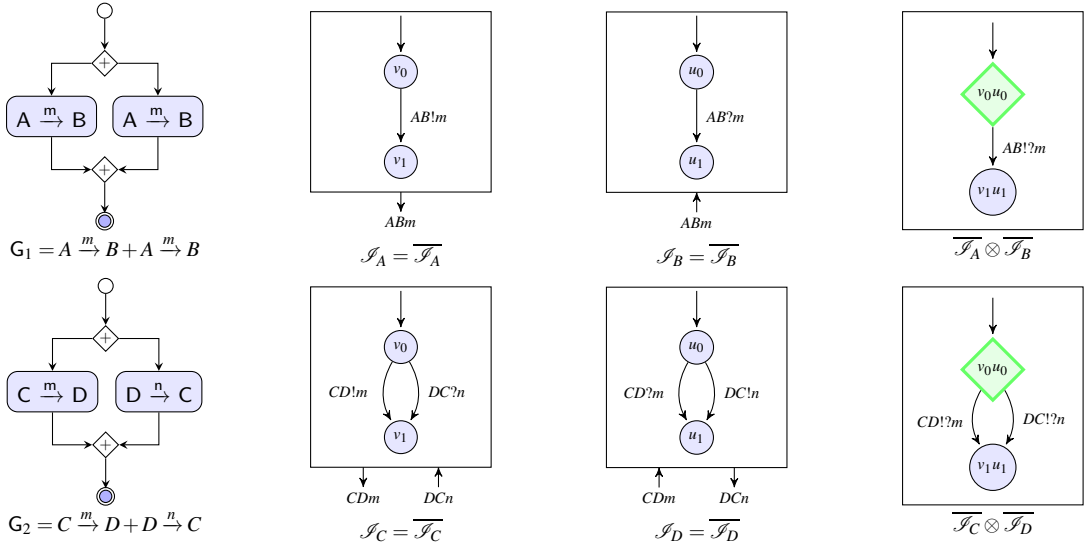
\begin{figure}[htbp]
\begin{minipage}{0.24\textwidth}
\centering
\begin{tikzpicture}[node distance=1cm,
    every node/.style={fill=white, font=\sffamily}, align=center]
  \node (0)             [start]              {};
  \node at (0, -0.7) (1)     [branch]          {\tiny $+$};
  \node (2)     [basic, below left of=1]          {\scriptsize A $\xrightarrow{\text{m}}$ B};
  \node (3)      [basic, below right of=1]   {\scriptsize A $\xrightarrow{\text{m}}$ B};
  \node (4)    [branch, below right of=2]              {\tiny $+$};
  \node at (0, -2.8) (5)    [end]              {};
  \draw[-{stealth}]      (0) -- (1);
  \draw[-{stealth}]     (1) -| (2);
  \draw[-{stealth}]      (1) -| (3);
  \draw[-{stealth}]      (2) |- (4);
  \draw[-{stealth}]      (3) |- (4);
  \draw[-{stealth}]      (4) -- (5);
\end{tikzpicture}
\centering
\\  \scriptsize $\textsf{G}_1=A\xrightarrow{m}B + A\xrightarrow{m}B$
\end{minipage}
\begin{minipage}{0.24\textwidth}
\begin{tikzpicture}[shorten >=1pt,node distance=1.2cm,>=stealth',auto,
                                    every state/.style={thin,fill=blue!10}]      
\draw (-1.2,-1.6) rectangle (1.2,0.8);
\draw  [->] (-0.0,-1.6)--(-0.0,-1.9);
\node at (-0.0,-2.0) {\tiny $ABm$};                                                             
\node[state,initial above,inner sep=2pt,minimum size=0pt]        (q_0)                           {\tiny $v_0$};  
\node[state,inner sep=2pt,minimum size=0pt]  (q_1)  [below of=q_0]     {\tiny $v_1$}; 
\path[->]       
             (q_0)     edge  []  node{\tiny $AB!m$}     (q_1);
\end{tikzpicture}
\centering
\\ \scriptsize $\mathcal{I}_{A}=\overline{\mathcal{I}_{A}}$
\end{minipage}
\begin{minipage}{0.24\textwidth}
\begin{tikzpicture}[shorten >=1pt,node distance=1.2cm,>=stealth',auto,
                                    every state/.style={thin,fill=blue!10}]      
\draw (-1.2,-1.6) rectangle (1.2,0.8);
\draw  [<-] (-0.0,-1.6)--(-0.0,-1.9);
\node at (-0.0,-2.0) {\tiny $ABm$};                                                             
\node[state,initial above,inner sep=2pt,minimum size=0pt]        (q_0)                           {\tiny $u_0$};  
\node[state,inner sep=2pt,minimum size=0pt]  (q_1)  [below of=q_0]     {\tiny $u_1$}; 
\path[->]       
             (q_0)     edge  []  node{\tiny $AB?m$}     (q_1);
\end{tikzpicture}
\centering
\\ \scriptsize $\mathcal{I}_{B}=\overline{\mathcal{I}_{B}}$
\end{minipage}
\begin{minipage}{0.24\textwidth}
\begin{tikzpicture}[shorten >=1pt,node distance=1.3cm,>=stealth',auto,
                                    every state/.style={thin,fill=blue!10}, dia/.style={diamond, draw=green!60, very thick, fill=green!10}]      
\draw (-1.2,-1.8) rectangle (1.2,1.1);

\node[dia,initial above,inner sep=2pt,minimum size=0pt]        (q_0)                           {\tiny $v_0u_0$};  
\node[state,inner sep=2pt,minimum size=0pt]  (q_1)  [below of=q_0]     {\tiny $v_1u_1$};  
\path[->]       
             (q_0)    edge []    node{\tiny $AB!?m$}     (q_1);
             		  
\end{tikzpicture}
\centering 
\\ \scriptsize $\overline{\mathcal{I}_A} \otimes \overline{\mathcal{I}_B}$
\end{minipage}\newline

\begin{minipage}{0.24\textwidth}
\centering
\begin{tikzpicture}[node distance=1cm,
    every node/.style={fill=white, font=\sffamily}, align=center]
  \node (0)             [start]              {};
  \node at (0, -0.7) (1)     [branch]          {\tiny $+$};
  \node (2)     [basic, below left of=1]          {\scriptsize C $\xrightarrow{\text{m}}$ D};
  \node (3)      [basic, below right of=1]   {\scriptsize D $\xrightarrow{\text{n}}$ C};
  \node (4)    [branch, below right of=2]              {\tiny $+$};
  \node at (0, -2.8) (5)    [end]              {};
  \draw[-{stealth}]      (0) -- (1);
  \draw[-{stealth}]     (1) -| (2);
  \draw[-{stealth}]      (1) -| (3);
  \draw[-{stealth}]      (2) |- (4);
  \draw[-{stealth}]      (3) |- (4);
  \draw[-{stealth}]      (4) -- (5);
\end{tikzpicture}
\centering
\\  \scriptsize $\textsf{G}_2=C\xrightarrow{m}D + D\xrightarrow{n}C$
\end{minipage}
\begin{minipage}{0.24\textwidth}
\begin{tikzpicture}[shorten >=1pt,node distance=1.2cm,>=stealth',auto,
                                    every state/.style={thin,fill=blue!10}]      
\draw (-1.2,-1.6) rectangle (1.2,0.8);
\draw  [->] (-0.5,-1.6)--(-0.5,-1.9);
\draw  [<-] (0.5,-1.6)--(0.5,-1.9);
\node at (-0.5,-2.0) {\tiny $CDm$};                                             \node at (0.5,-2.0) {\tiny $DCn$};                                                             
\node at (-0.7,-0.6) {\tiny $CD!m$};                                                             
\node[state,initial above,inner sep=2pt,minimum size=0pt]        (q_0)                           {\tiny $v_0$};  
\node[state,inner sep=2pt,minimum size=0pt]  (q_1)  [below of=q_0]     {\tiny $v_1$}; 
\path[->]       
             (q_0)   edge  [bend right]  node{}     (q_1)  
                     edge  [bend left]  node{\tiny $DC?n$}     (q_1);
\end{tikzpicture}
\centering
\\ \scriptsize $\mathcal{I}_{C}=\overline{\mathcal{I}_{C}}$
\end{minipage}
\begin{minipage}{0.24\textwidth}
\begin{tikzpicture}[shorten >=1pt,node distance=1.2cm,>=stealth',auto,
                                    every state/.style={thin,fill=blue!10}]      
\draw (-1.2,-1.6) rectangle (1.2,0.8);
\draw  [<-] (-0.5,-1.6)--(-0.5,-1.9);
\draw  [->] (0.5,-1.6)--(0.5,-1.9);
\node at (-0.5,-2.0) {\tiny $CDm$};                                             \node at (0.5,-2.0) {\tiny $DCn$};                                           
\node at (-0.7,-0.6) {\tiny $CD?m$};                                                             
\node[state,initial above,inner sep=2pt,minimum size=0pt]        (q_0)                           {\tiny $u_0$};  
\node[state,inner sep=2pt,minimum size=0pt]  (q_1)  [below of=q_0]     {\tiny $u_1$}; 
\path[->]       
             (q_0)  edge  [bend right]  node{}     (q_1)   
                    edge  [bend left]  node{\tiny $DC!n$}     (q_1);
\end{tikzpicture}
\centering
\\ \scriptsize $\mathcal{I}_{D}=\overline{\mathcal{I}_{D}}$
\end{minipage}
\begin{minipage}{0.24\textwidth}
\begin{tikzpicture}[shorten >=1pt,node distance=1.3cm,>=stealth',auto,
                                    every state/.style={thin,fill=blue!10}, dia/.style={diamond, draw=green!60, very thick, fill=green!10}]      
\draw (-1.2,-1.8) rectangle (1.2,1.1);
\node at (-0.7,-0.7) {\tiny $CD!?m$};                                                             

\node[dia,initial above,inner sep=2pt,minimum size=0pt]        (q_0)                           {\tiny $v_0u_0$};  
\node[state,inner sep=2pt,minimum size=0pt]  (q_1)  [below of=q_0]     {\tiny $v_1u_1$};  
\path[->]       
             (q_0)   edge [bend right]    node{}     (q_1) 
                     edge [bend left]    node{\tiny $DC!?n$}     (q_1);
             		  
\end{tikzpicture}
\centering 
\\ \scriptsize $\overline{\mathcal{I}_C} \otimes \overline{\mathcal{I}_D}$
\end{minipage}
\caption{Branching error states in $\otimes_{A\in \mathcal{P}}(\textsf{G}_{1\downarrow A})$  and $\otimes_{C\in \mathcal{P}}(\textsf{G}_{2\downarrow C})$}\label{fig:vbranShowErrorStateNotEnough}
\end{figure}

Branching error states capture the non well-formedness of branching g-choreographies that can not be detected by error states. According to Definition~\ref{branchingDefErrorState}, we summarise four classes of branching error states. Let $\textsf{G}$ be a branching g-choreography. 

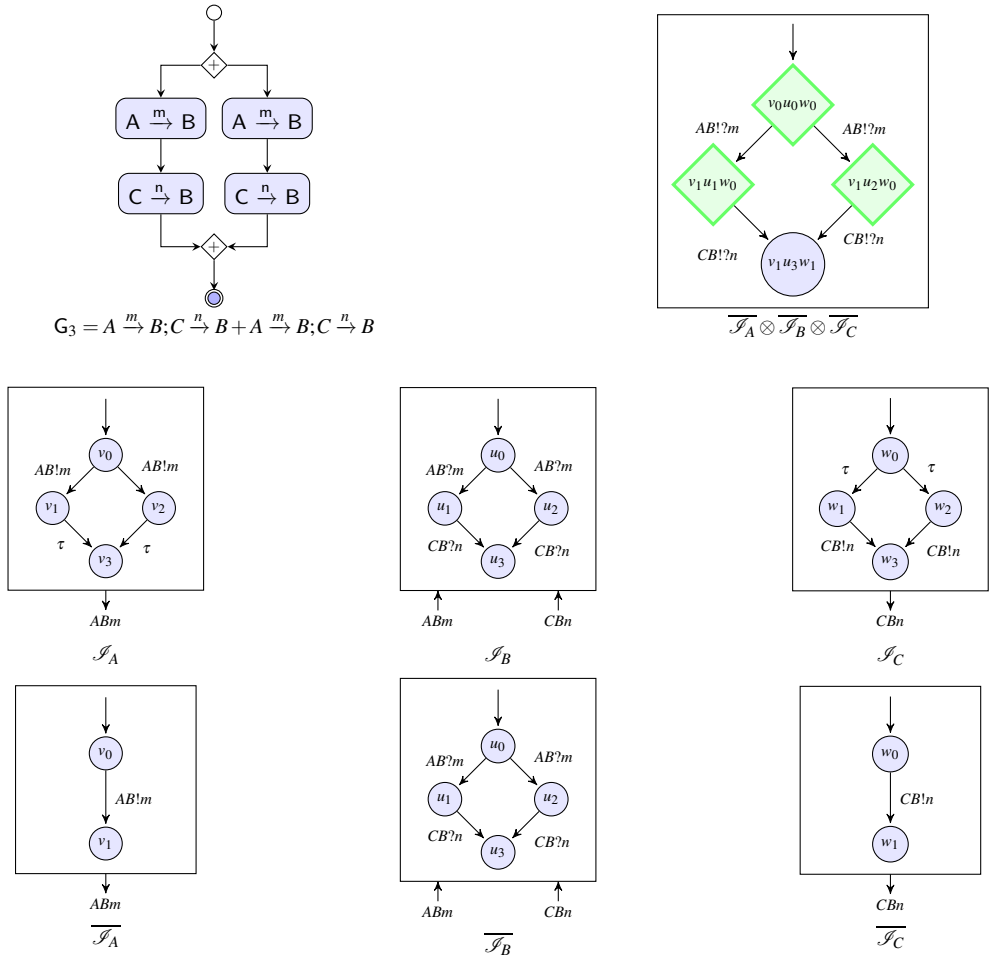
\begin{figure}[h]
\begin{minipage}{0.5\textwidth}
\centering
\begin{tikzpicture}[node distance=1cm,
    every node/.style={fill=white, font=\sffamily}, align=center]
  \node (0)             [start]              {};
  \node at (0, -0.7) (1)     [branch]          {\tiny $+$};
  \node (2)     [basic, below left of=1]          {\scriptsize A $\xrightarrow{\text{m}}$ B};
  \node (3)     [basic, below of=2]          {\scriptsize C $\xrightarrow{\text{n}}$ B};
  \node (4)      [basic, below right of=1]   {\scriptsize A $\xrightarrow{\text{m}}$ B};
  \node (5)     [basic, below of=4]          {\scriptsize C $\xrightarrow{\text{n}}$ B};
  \node (6)    [branch, below right of=3]              {\tiny $+$};
  \node at (0, -3.8) (7)    [end]              {};
  \draw[-{stealth}]      (0) -- (1);
  \draw[-{stealth}]      (1) -| (2);
  \draw[-{stealth}]      (2) -- (3);
  \draw[-{stealth}]      (1) -| (4);
  \draw[-{stealth}]      (4) -- (5);
  \draw[-{stealth}]      (3) |- (6);
  \draw[-{stealth}]      (5) |- (6);
  \draw[-{stealth}]      (6) -- (7);
\end{tikzpicture}
\centering
\\  \scriptsize $\textsf{G}_3=A\xrightarrow{m}B;C\xrightarrow{n}B+A\xrightarrow{m}B;C\xrightarrow{n}B$
\end{minipage}
\begin{minipage}{0.45\textwidth}
\begin{tikzpicture}[shorten >=1pt,node distance=1.5cm,>=stealth',auto,
                                    every state/.style={thin,fill=blue!10}, dia/.style={diamond, draw=green!60, very thick, fill=green!10}]      
\draw (-1.8,-2.7) rectangle (1.8,1.2);
\node at (-1.0,-0.3) {\tiny $AB!?m$};                                          
\node at (-1.0,-2.0) {\tiny $CB!?n$};                                    
\node[dia,initial above,inner sep=2pt,minimum size=0pt]        (q_0)                           {\tiny $v_0u_0w_0$};  
\node[dia,inner sep=2pt,minimum size=0pt]  (q_1)  [below left of=q_0]     {\tiny $v_1u_1w_0$}; 
\node[dia,inner sep=2pt,minimum size=0pt]  (q_2)  [below right of=q_0]     {\tiny $v_1u_2w_0$}; 
\node[state,inner sep=2pt,minimum size=0pt]  (q_3)  [below right of=q_1]     {\tiny $v_1u_3w_1$}; 
\path[->]       
             (q_0)    edge    node{}     (q_1)            		 
             (q_0)    edge    node{\tiny $AB!?m$}     (q_2)
             (q_1)    edge []   node{}     (q_3) 
             (q_2)    edge []   node{\tiny $CB!?n$}     (q_3);
\end{tikzpicture}
\centering 
\\ \scriptsize $\overline{\mathcal{I}_A} \otimes \overline{\mathcal{I}_B} \otimes \overline{\mathcal{I}_C}$
\end{minipage}\newline\newline

\begin{minipage}{0.32\textwidth}
\begin{tikzpicture}[shorten >=1pt,node distance=1.0cm,>=stealth',auto,
                                    every state/.style={thin,fill=blue!10}]
\draw (-1.3,-1.8) rectangle (1.3,0.9);
\node at (-0.7,-0.2) {\tiny $AB!m$};
\node at (-0.6,-1.2) {\tiny $\tau$};
\draw  [->] (-0.0,-1.8)--(-0.0,-2.1);
\node at (-0.0,-2.2) {\tiny $ABm$};
\node[state,initial above,inner sep=2pt,minimum size=0pt]        (u_0)                           {\tiny $v_0$};  
\node[state,inner sep=2pt,minimum size=0pt]  (u_1)  [below left of=u_0]     {\tiny $v_1$}; 
\node[state,inner sep=2pt,minimum size=0pt]  (u_2)  [below right of=u_0]     {\tiny $v_2$};
\node[state,inner sep=2pt,minimum size=0pt]  (u_3)  [below  right of=u_1]     {\tiny $v_3$};
\path[->]       
             (u_0)     edge    node{}     (u_1)
                       edge    node{\tiny $AB!m$}     (u_2) 
             (u_1)     edge    node{\tiny }     (u_3)    
             (u_2)     edge    node{\tiny $\tau$}     (u_3);
\end{tikzpicture}
\centering
\\ \scriptsize $\mathcal{I}_{A}$
\end{minipage}
\begin{minipage}{0.32\textwidth}
\begin{tikzpicture}[shorten >=1pt,node distance=1.0cm,>=stealth',auto,
                                    every state/.style={thin,fill=blue!10}]
\draw (-1.3,-1.8) rectangle (1.3,0.9);
\node at (-0.7,-0.2) {\tiny $AB?m$};
\node at (-0.7,-1.2) {\tiny $CB?n$};
\draw  [<-] (-0.8,-1.8)--(-0.8,-2.1);
\draw  [<-] (0.8,-1.8)--(0.8,-2.1);
\node at (-0.8,-2.2) {\tiny $ABm$};
\node at (0.8,-2.2) {\tiny $CBn$};
\node[state,initial above,inner sep=2pt,minimum size=0pt]        (u_0)                           {\tiny $u_0$};  
\node[state,inner sep=2pt,minimum size=0pt]  (u_1)  [below left of=u_0]     {\tiny $u_1$}; 
\node[state,inner sep=2pt,minimum size=0pt]  (u_2)  [below right of=u_0]     {\tiny $u_2$};
\node[state,inner sep=2pt,minimum size=0pt]  (u_3)  [below  right of=u_1]     {\tiny $u_3$};
\path[->]       
             (u_0)     edge    node{}     (u_1)
                       edge    node{\tiny $AB?m$}     (u_2) 
             (u_1)     edge    node{\tiny }     (u_3)    
             (u_2)     edge    node{\tiny $CB?n$}     (u_3);
\end{tikzpicture}
\centering
\\ \scriptsize $\mathcal{I}_{B}$
\end{minipage}
\begin{minipage}{0.32\textwidth}
\begin{tikzpicture}[shorten >=1pt,node distance=1.0cm,>=stealth',auto,
                                    every state/.style={thin,fill=blue!10}]
\draw (-1.3,-1.8) rectangle (1.3,0.9);
\node at (-0.6,-0.2) {\tiny $\tau$};
\node at (-0.7,-1.2) {\tiny $CB!n$};
\draw  [->] (-0.0,-1.8)--(-0.0,-2.1);
\node at (0.0,-2.2) {\tiny $CBn$};
\node[state,initial above,inner sep=2pt,minimum size=0pt]        (u_0)                           {\tiny $w_0$};  
\node[state,inner sep=2pt,minimum size=0pt]  (u_1)  [below left of=u_0]     {\tiny $w_1$}; 
\node[state,inner sep=2pt,minimum size=0pt]  (u_2)  [below right of=u_0]     {\tiny $w_2$};
\node[state,inner sep=2pt,minimum size=0pt]  (u_3)  [below  right of=u_1]     {\tiny $w_3$};
\path[->]       
             (u_0)     edge    node{}     (u_1)
                       edge    node{\tiny $\tau$}     (u_2) 
             (u_1)     edge    node{\tiny }     (u_3)    
             (u_2)     edge    node{\tiny $CB!n$}     (u_3);
\end{tikzpicture}
\centering
\\ \scriptsize $\mathcal{I}_{C}$
\end{minipage}\newline

\begin{minipage}{0.32\textwidth}
\begin{tikzpicture}[shorten >=1pt,node distance=1.2cm,>=stealth',auto,
                                    every state/.style={thin,fill=blue!10}]      
\draw (-1.2,-1.6) rectangle (1.2,0.9);
\draw  [->] (-0.0,-1.6)--(-0.0,-1.9);
\node at (-0.0,-2.0) {\tiny $ABm$};                                     
\node[state,initial above,inner sep=2pt,minimum size=0pt]        (q_0)                           {\tiny $v_0$};  
\node[state,inner sep=2pt,minimum size=0pt]  (q_1)  [below of=q_0]     {\tiny $v_1$}; 
\path[->]       
             (q_0)     edge  []  node{\tiny $AB!m$}     (q_1);
\end{tikzpicture}
\centering
\\ \scriptsize $\overline{\mathcal{I}_{A}}$
\end{minipage}
\begin{minipage}{0.32\textwidth}
\begin{tikzpicture}[shorten >=1pt,node distance=1.0cm,>=stealth',auto,
                                    every state/.style={thin,fill=blue!10}]
\draw (-1.3,-1.8) rectangle (1.3,0.9);
\node at (-0.7,-0.2) {\tiny $AB?m$};
\node at (-0.7,-1.2) {\tiny $CB?n$};
\draw  [<-] (-0.8,-1.8)--(-0.8,-2.1);
\draw  [<-] (0.8,-1.8)--(0.8,-2.1);
\node at (-0.8,-2.2) {\tiny $ABm$};
\node at (0.8,-2.2) {\tiny $CBn$};
\node[state,initial above,inner sep=2pt,minimum size=0pt]        (u_0)                           {\tiny $u_0$};  
\node[state,inner sep=2pt,minimum size=0pt]  (u_1)  [below left of=u_0]     {\tiny $u_1$}; 
\node[state,inner sep=2pt,minimum size=0pt]  (u_2)  [below right of=u_0]     {\tiny $u_2$};
\node[state,inner sep=2pt,minimum size=0pt]  (u_3)  [below  right of=u_1]     {\tiny $u_3$};
\path[->]       
             (u_0)     edge    node{}     (u_1)
                       edge    node{\tiny $AB?m$}     (u_2) 
             (u_1)     edge    node{\tiny }     (u_3)    
             (u_2)     edge    node{\tiny $CB?n$}     (u_3);
\end{tikzpicture}
\centering
\\ \scriptsize $\overline{\mathcal{I}_{B}}$
\end{minipage}
\begin{minipage}{0.32\textwidth}
\begin{tikzpicture}[shorten >=1pt,node distance=1.2cm,>=stealth',auto,
                                    every state/.style={thin,fill=blue!10}]      
\draw (-1.2,-1.6) rectangle (1.2,0.9);
\draw  [->] (-0.0,-1.6)--(-0.0,-1.9);
\node at (-0.0,-2.0) {\tiny $CBn$};                                                             
\node[state,initial above,inner sep=2pt,minimum size=0pt]        (q_0)                           {\tiny $w_0$};  
\node[state,inner sep=2pt,minimum size=0pt]  (q_1)  [below of=q_0]     {\tiny $w_1$}; 
\path[->]       
             (q_0)     edge []   node{\tiny $CB!n$}     (q_1);
\end{tikzpicture}
\centering
\\ \scriptsize $\overline{\mathcal{I}_{C}}$
\end{minipage}
\caption{Branching error states in $\otimes_{A\in \mathcal{P}}(\textsf{G}_{3\downarrow A})$}\label{fig:vbranShowErrorStateNotEnough''}
\end{figure}

$(1)$ If the initial state of $\otimes_{A\in\mathcal{P}}{(\textsf{G'}+\textsf{G''})}_{\downarrow A}$ just has only one transition, then the initial state is a branching error state. For instance, in Figure~\ref{fig:vbranShowErrorStateNotEnough}, we are able to see the initial state $v_0u_0$ denoted by a green diamond of $\overline{\mathcal{I}_A} \otimes \overline{\mathcal{I}_B}$ is a branching error state (by~\eqref{brdef:1} in Definition~\ref{branchingDefErrorState}) since there is only transitions from $v_0u_0$ to $v_1u_1$.
$(2)$ If the initial state of $\otimes_{A\in\mathcal{P}}{(\textsf{G'}+\textsf{G''})}_{\downarrow A}$ has two transitions which have different subjects (sender) in transition actions, then, we define this initial state is a branching error state. For instance, in Figure~\ref{fig:vbranShowErrorStateNotEnough}, we are able to see the initial state $v_0u_0$ denoted by a green diamond of $\overline{\mathcal{I}_C} \otimes \overline{\mathcal{I}_D}$ is a branching error state (by~\eqref{brdef:2} in Definition~\ref{branchingDefErrorState}) since there exist two transitions from $v_0u_0$ to $v_1u_1$, where their transition actions are $CD!?m$, $DC!?n$ respectively representing two different senders $C$ and $D$ at $v_0u_0$.
$(3)$ If the initial state of $\otimes_{A\in\mathcal{P}}{(\textsf{G'}+\textsf{G''})}_{\downarrow A}$ has two transitions which have the same transition actions, then, we define this initial state is a branching error state. For instance, in Figure~\ref{fig:vbranShowErrorStateNotEnough''}, we are able to see the initial state $v_0u_0w_0$ denoted by a green diamond of $\overline{\mathcal{I}_A} \otimes \overline{\mathcal{I}_B} \otimes \overline{\mathcal{I}_C} $ is a branching error state (by~\eqref{brdef:2} in Definition~\ref{branchingDefErrorState}) since there are two transitions from $v_0u_0w_0$ to have the same actions $AB!?m$.  
$(4)$ If there exist two states located in the execution paths from the initial state of $\otimes_{A\in\mathcal{P}}{(\textsf{G'}+\textsf{G''})}_{\downarrow A}$, and when each of these two states has a transition which has the same transition action, while the subject of the action is not in the set of the subjects and objects of the strings from the initial states to these two states, then, we define these two states are branching error states. For instance, in Figure~\ref{fig:vbranShowErrorStateNotEnough''}, we are able to see the states $v_1u_1w_0$ and $v_1u_2w_0$ denoted by green diamonds of $\overline{\mathcal{I}_A} \otimes \overline{\mathcal{I}_B}\otimes \overline{\mathcal{I}_C}$ are branching error states (by~\eqref{brdef:3} in Definition~\ref{branchingDefErrorState}) since there are two transitions $v_1u_1w_0\xrightarrow{CB!?n}v_1u_3w_1$ and $v_1u_2w_0\xrightarrow{CB!?n}v_1u_3w_1$, where they have the same transition action and $C$ is not in $\textit{sobj}(AB!?m)$. 

\subsection{Well-formedness and non-existence of (parallel, branching) error states}\label{secMainTheory}
We have introduced projection from g-choreographies to GIA, the removability of $\tau$-transitions, branching error states and parallel error states. Next, we write the main theorem below.
\begin{theorem}[]\label{maintheorem}
A g-choreography $\textsf{G}$ is well-formed iff there are neither error states nor parallel, branching error states in $\otimes_{A \in \mathcal{P}}(\overline{{\textsf{G}_{\downarrow A}}}),$
where $\overline{{\textsf{G}_{\downarrow A}}}$ is the GIA without removable $\tau$-transition constructed from $\textsf{G}_{\downarrow A}$.
\end{theorem} 

The theorem~\ref{maintheorem} shows that a g-choreography (Definition~\ref{defGG}) is well-formed (Definition~\ref{semanticOfGGSimple}) if, and only if the $\otimes$-product (Definition~\ref{productOfAIA}) of the set of projections (Definition~\ref{defIAFromGG})  without $\tau$ (Definition~\ref{remtauIA},~\ref{IAwithoutTau}) does not have error states (Definition~\ref{defErrorState}), parallel error states(Definition~\ref{parDefErrorState}) and 
branching error states (Definition~\ref{branchingDefErrorState}).

\section{Conclusion}
We established a new way to check well-formedness based on an extension of
interface automata.
We adopted a variant of the semantics of g-choreograhies
presented in~\cite{gt16}.
Our semantics relaxes some conditions on well-formedness (more
precisely the conditions for sequential, parallel, and branching
compositions are stricter in~\cite{gt16} than here).
In this paper we did not consider iterative g-choreograhies; however,
extending our results to this case is conceptually straightforward as
loops can be dealt with as done eg in~\cite{fmt18}.
The main limitation of our notion of well-formedness is in the treatment of
well-branchedness that here we treat \quo{syntactically}.
For instance, in $\textsf{G}_1$ of
Figure~\ref{fig:vbranShowErrorStateNotEnough}, participants $A$ and
$B$ are neither active nor passive participants according to our since
$A$ and $B$ behave exactly same in both branches making the choice non
well-branched.
In fact, to identify active and passive participants our semantics relies
on divergence points imposed by the syntax of g-choreograhies instead
of using the (semantic) concept of prefix-maps.
This latter concept allows one to identify, for each participant, where
(if at all) the participant becomes aware of the choice.
Hence, using prefix maps the choice $\textsf{G}_1$ of
Figure~\ref{fig:vbranShowErrorStateNotEnough} is well-branched since
both $A$ and $B$ behaves uniformly in the branches ($A$ and $B$ are
both passive and therefore unaware of the choice which is indeed not a
\quo{semantic} choice).
Here we opted for simplicity; our results can be casted in the more
general setting at the cost of increasing the technical complexity.

As a matter of fact, finding removable $\tau$ and obtaining refined projections has an exponential theoretical complexity.
This is mainly due to the minimisation of finite state machines. We
believe that in practice this is not a great problem since in practice
minimisation is rather effective. We are developing a prototype tool
(\url{https://github.com/haomoons/GIAGG}) to conduct experiments and
measure how our approach performs in practice available at to support
our theory.




\bibliographystyle{eptcs}
\bibliography{general}
\end{document}
\grid